\documentclass[journal]{IEEEtran}
\usepackage{cite}
\usepackage{amsmath,amssymb,amsfonts}
\usepackage{algorithmic}
\usepackage{graphicx}
\usepackage{textcomp}
\usepackage{xcolor}
\usepackage{stfloats}
\usepackage{verbatim}
\ifCLASSOPTIONcompsoc
\usepackage[caption=false, font=normalsize, labelfont=sf, textfont=sf]{subfig}
\else
\usepackage[caption=false, font=footnotesize]{subfig}
\fi

\usepackage{cuted}
\usepackage{url}								
\usepackage{siunitx}							
\usepackage{booktabs}							
\usepackage{tabularx}
\newcolumntype{Y}{>{\centering\arraybackslash}X} 

\usepackage{makecell}							
\usepackage{multirow}							
\usepackage[display-foreign=false]{acro}			
\usepackage{gensymb}
\acsetup{single}								

\usepackage{tikz}								

\usepackage{soul}
\usepackage[most]{tcolorbox}

\DeclareAcronym{LCMP}{
	short = LCMP,
	long = linear constrained minimum power 
}

\begin{document}
	
	\title{Multi-Objective Adaptive Beamforming Using Partial Knowledge of Dynamic Dielectric Media for Non-Invasive Microwave Hyperthermia}
	
	\author{ Ahona Bhattacharyya,~\IEEEmembership{Graduate~Student~Member,~IEEE,} 
		Tessa A. Haldes,~\IEEEmembership{Graduate~Student~Member,~IEEE},\\
		Jeffrey A. Nanzer,~\IEEEmembership{Senior Member,~IEEE},
		and Susan C. Hagness~\IEEEmembership{Fellow,~IEEE}
		
		\thanks{This work was supported in part by the Defense Advanced Research Projects Agency under grant \#HR00112010015, the Office of Naval Research under grant \#N00014-20-1-2389, and the National Science Foundation under grant \#1751655. \textit{(Corresponding author: Ahona Bhattacharyya)}}
		\thanks{	
			A. Bhattacharyya and J. A. Nanzer are with the Department of Electrical and Computer Engineering, Michigan State University, East Lansing, MI 48824 USA (email: bhatta67@msu.edu, nanzer@msu.edu). T. Haldes and S. C. Hagness are with the Department of Electrical and Computer Engineering, the University of Wisconsin-Madison, Madison, WI 53706 USA (email: haldes@wisc.edu, susan.hagness@wisc.edu).}
		
	}
	
	\maketitle
	\setlength{\stripsep}{-5.5em}
	\begin{strip}
		\noindent \textbf{\textit{Abstract}} We investigate multi-objective adaptive beamformer design strategies for non-invasive microwave hyperthermia. Our focus is to address the challenges of maintaining focused power deposition in desired locations while reducing unwanted heating elsewhere under conditions of changing dielectric properties. The process of heating the media causes changes in the dielectric properties of the media, which can degrade the effectiveness of the beamformers with static weights. Typical hyperthermic beamformer designs calculate antenna beamforming weights using patient-specific high resolution dielectric maps obtained by MRI or microwave tomography; however, this process is time consuming and difficult to perform in real-time. In this work, we explore the efficacy of microwave hyperthermia in various inhomogeneous media under changing dielectric conditions, with the goal of informing the design of future adaptive real-time microwave hyperthermia techniques. We aim to achieve cell apoptosis by obtaining temperatures of $\sim$ \SI{45}{\degreeCelsius} through selective absorption of electromagnetic waves focusing at a \SI{2.5}{\giga\hertz} carrier frequency with little to no knowledge of the changes in the dielectric media and simultaneously place nulls to avoid unwanted heating outside of the treatment zone. We investigate the effectiveness of the linear constrained minimum power (LCMP) algorithm for near-field multi-objective beamforming and examine the power density obtained from finite-difference time-domain (FDTD) simulations on simple analytical models and anatomically realistic numerical breast phantoms. To gain a comprehensive knowledge of the efficacy of the beamformer we evaluate the resulting thermal maps of the models in simple homogeneous cases, heterogeneous cases and MRI-derived phantom breast models.
		\\
		\\
		\noindent \textbf{\textit{Keywords}} --- Microwave Hyperthermia,  multi-objective beamforming, near-field beamforming, synchronization
		\vspace{7em}
	\end{strip}
	
	\acresetall 
	
	\vspace{50em}
	\section{Introduction}
	\label{sec:intro}
	
	Application of non-invasive microwave hyperthermia to treat sub-surface level cancerous tumors has seen a rise in interest in the recent years. The idea of using microwave hyperthermia as an aid to more traditional methods of treatment of various types of cancer has been explored for a few decades. The objective of hyperthermic treatment is to heat the cancerous cells to a temperature that makes it more susceptible to chemotherapy or radiation therapy, or to apoptosis temperatures between \SI{42}{\degreeCelsius} to \SI{50}{\degreeCelsius} to trigger cell death. The authors in~\cite{valdagni1994report} compared clinical trial results of noncancerous cell survival and long term damage to patients with inoperable metastatic neck lymph nodes. It was found that patients treated with radiation therapy plus localized hyperthermia showed improved survival compared to only conventional radiation therapy alone. Similar cases of the use of hyperthermia as a supplemental treatment for treating cancer are discussed in~\cite{overgaard1995randomised, kapp1996efficacy}.
	
	Some of the earliest reported studies on non-invasive microwave hyperthermia are reported in \cite{1987_center_emdistribution, 1992_offcenter_emdistribution}. In \cite{1991_computer_fdtd} the authors discuss the use of the standard finite-difference time-domain (FDTD) algorithm~\cite{1987_fdtd} to solve for the 2D electric and magnetic components for microwave hyperthermia. 
	The FDTD method's detailed near-field analysis capabilities, including the evaluation of surface currents, energy deposition, and field concentrations, make it a valuable asset for researchers in antenna design, microwave engineering, and bio-electromagnetics. In~\cite{1996_fdtd_verification}, non-invasive microwave hyperthermia is studied in simulated power deposition measurements of a MRI-derived model of a brain with a tumour using a 3D-FDTD algorithm. 
	Other studies include~\cite{bernardi2003specific}, where the patient was kept in the far-field to the sources to evaluate the thermal changes of the tissue, as well as~\cite{gosalia2004thermal} where an FDTD model was used to evaluate temperature changes due to an embedded IC chip of a retinal prosthesis used to restore partial vision.
	
	Significant progress has been made on minimally invasive hyperthermia treatment for breast cancer over the last few decades~\cite{fenn1999adaptive, 2002_ajfenn_microwave_phased_array_Gardner2002, 2004_ajfenn_phased_array_Vargas2004, 2010_ajfenn_phased_array_Dooley2010}. These clinical studies explored adaptive microwave phased arrays with invasive feedback, employing breast compression to stabilize tissue and enhance microwave energy delivery. In~\cite{mohtashami2020ex,evans2021feasibility}, minimally invasive ablation antennas were designed and experimentally validated using continuous and pulsed waveforms. 
	In parallel, theoretical work has examined alternative non-invasive focusing strategies like time reversal~\cite{guo2005time, kosmas2007computational}, deformable mirrors~\cite{arunachalam2007computational} and transmit beamforming~\cite{converse2004ultrawide, converse2006computational} using computational simulations on 2D breast phantoms. The success of the designed beamformers in these works relies on prior knowledge of the breast tissue properties obtained from MRI derived models. The authors in~\cite{gouzouasis2007fdtd} studied the use of FDTD simulations to focus energy in numerical phantoms of the human head and in~\cite{karanasiou2008development} they experimentally validate their theoretical findings. In~\cite{burfeindt2011microwave} a patient specific model of human brain was used to design a microwave array to treat pediatric brain tumors. 
	The two most commonly researched applications of the non-invasive tumors treatments are in human head or for breast cancer models, which can be built from available MRI-derived dielectric models.
	Patient-specific beamformer designs can be simulated from these readily available models, as demonstrated in~\cite{zastrow20103d}, where design consisted of a finite-impulse response (FIR) on each transmit channel to manipulate the amplitude and phase of the transmit signal to focus energy at the target region. 
	Some non-invasive hyperthermia treatment plans that have been experimentally demonstrated include use of a water-loaded cavity-backed phased array at \SI{434}{\mega\hertz}~\cite{baskaran2020design} and particle swarm optimization~\cite{nguyen20173}. 
	
	The biggest challenge of designing a beamforming system for non-invasive microwave hyperthermia is to deliver heat precisely to the tumor and maintaining it, while avoiding heating of healthy tissue, despite inherent inaccuracies in the dielectric phantom models. Furthermore, the dielectric properties of the media are generally temperature dependent, such that during the hyperthermia process the propagation environment will change, rendering the beamforming weights inaccurate, which can lead to reduced focusing at the intended target area as well as unwanted increased heating in healthy areas. The non-invasive beamformers designed by the authors in the papers mentioned above were non-adaptive to dielectric changes that may occur in the tissue properties due to heating. While MRIs provide predominantly accurate dielectric maps and can be used to create the initial propagation models for testing hyperthermia, they are time consuming and costly, rendering them unsuitable for updating the beamformers to real time changes during the treatment. This has motivated research on using other methods of dielectric imaging like Confocal Microwave Imaging algorithms~\cite{o2016estimating}, radar based tissue sensing systems~\cite{garrett2015average,kurrant2015evaluation,lavoie2016estimating}, thermoacoustics~\cite{li2021preclinical}, electrical impedance tomography (EIT) \cite{zhou2023fft}, and estimation of average dielectric properties~\cite{bourqui2016system,haldes2024estimating}.         
	In addition to the challenge of changing dielectric properties, the beamformer must still be able to initially focus energy at intended areas while minimizing power deposition in healthy tissue. Some of this heating is challenging to avoid due to the high water content and higher conductivity of tissues like fibroglandular tissue or cerebrospinal fluid that causes increased power deposition.
	In~\cite{2011_Zastrow_time_multiplexing} the authors used a time-multiplexing scheme to switch through multiple layers of beamformers to non-invasively maintain focusing at a target in a 3D virtual patient model of a head while reducing unwanted auxiliary heating. In~\cite{baskaran2021multiobjective}, a multi-objective genetic algorithm was used to simultaneously maintain focus and reduce heating elsewhere. 
	\begin{figure}
		\centering
		\includegraphics[width=1\linewidth]{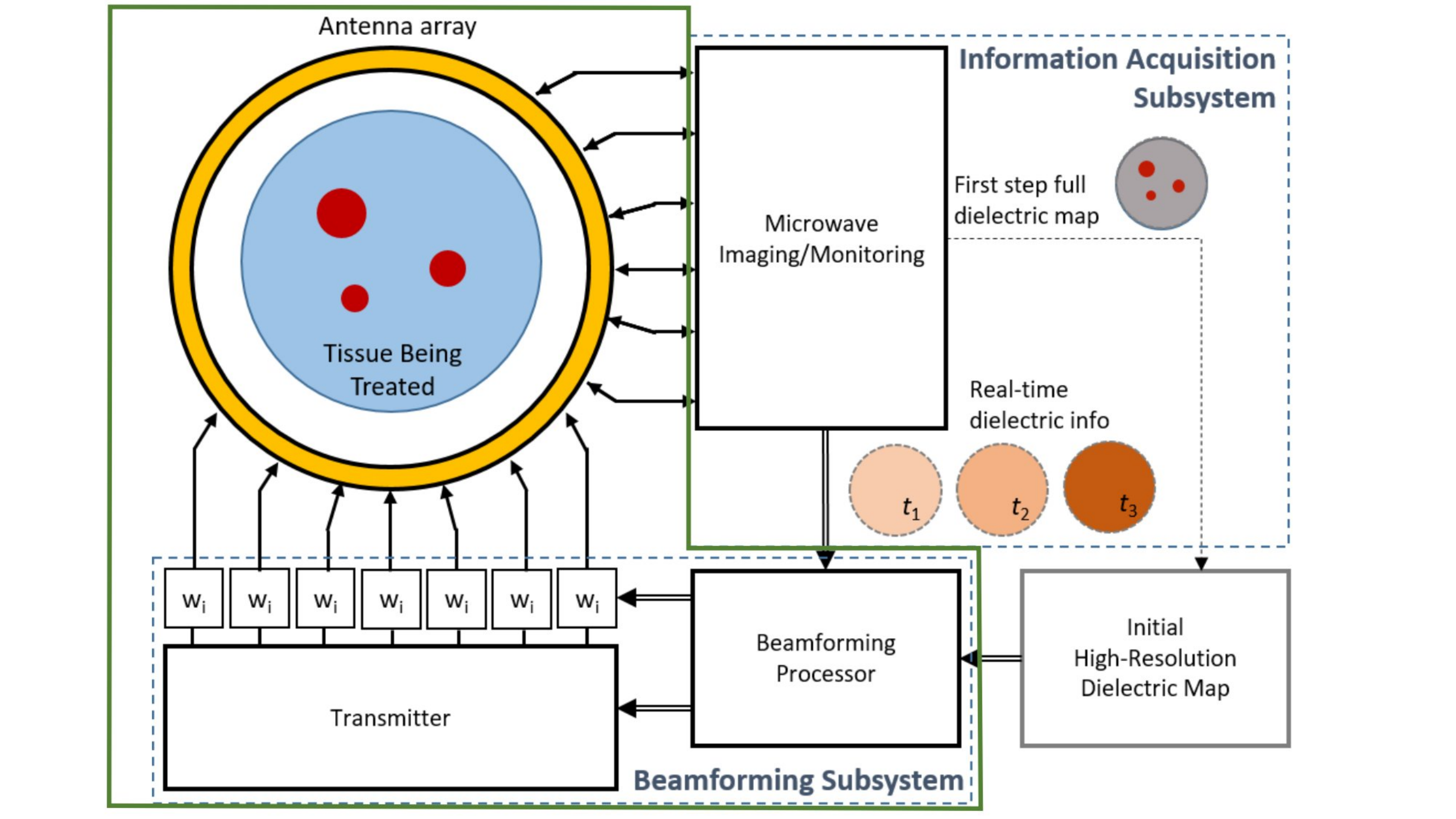}
		\caption{Conceptual diagram of the knowledge enhanced adaptive near-field multi-objective beamforming system for microwave hyperthermia with primary focus on designing the beamforming system and assuming that the imaging system is available.}
		\label{fig:overview} 
		\vspace{-\baselineskip}
	\end{figure}
	
	In this work we explore critical aspects of the design of a completely non-invasive multi-objective beamformer for hyperthermia which can adapt its beamforming weights according to the changes in the environment as reported by a real-time dielectric imaging system. The system concept is shown in Fig. \ref{fig:overview}. The advantage of this approach lies in the ability to operate using partial or spatially averaged dielectric change knowledge \cite{bhattacharyya2021selective, haldes2021analysis, bhattacharyya2022analysis, haldes2022adaptive3d, haldes2023estimate}, unlike competing techniques that necessitate ideal knowledge~\cite{zastrow20103d, 2011_Zastrow_time_multiplexing, baskaran2021multiobjective}.
	We focus in particular on the design of the beamformer and the impact of changing dielectric properties due to heating of the media, and the subsequent impact on beamforming efficacy. We furthermore explore the capability of an adaptive beamforming system to compensate for changing dielectric properties based on an estimate of the bulk dielectric changes in the environment. Starting with a high-resolution dielectric map obtained from MRI or other imaging approach, the proposed approach relies only on estimates of the change of the average dielectric properties of the media, representing information that can be more readily obtained by imaging methods.
	We base our study on designing beamformers for treating human breast cancer tumors. We used different analytical tissue models with varying complexity as test beds, starting with simple homogeneous models before moving to complex MRI-derived human breast phantoms, to test the efficacy of the beamformers. Computational methods are used to design and assess the thermal therapy provided by single objective beamformers across the different testbed setups and the success of the beamformer when only an average of the dielectric map is available. The study also involved testing the feasibility of using a multi-objective beamformer when there is unwanted heating in healthy tissue regions. Thus, this paper details crucial steps involved in successfully designing an adaptive multi-objective beamformer for non-invasive microwave hyperthermia applications.    
	
	\section{Methods}
	\label{sec:modeling}
	
	In this paper we aim to design a fully non-invasive beamformer for microwave hyperthermia treatment that can adapt to changes in the dielectric environment. 
	The simulations in this study are inspired by breast tissue hyperthermia treatment, thus the tissue properties, target locations, and antenna placement are chosen to mimic hyperthermia administered on breast tissue. The target was selected to be at least \SI{2}{\centi\meter} from the tissue boundary, to more closely represent realistic tumor locations. The thermal therapy efficacy was evaluated using coupled electromagnetic and thermal numerical simulations.
	We define a treatment region and a target region as two distinct tissue areas of interest. The target region is defined as the \SI{1}{\centi\meter} radial region encircling the target cell. This region represents the cancerous tissue. The treatment region is defined as the \SI{2}{\centi\meter} radial region encircling the target cell, including the target region. This region represents an additional margin of tissue that is relevant in microwave hyperthermia to ensure adequate thermal dose. Hereafter, in this paper the target region specifically refers to the tissue area encompassed by a \SI{1}{\centi\meter} radius and the treatment region is encompassed by a \SI{2}{\centi\meter} radius.
	We then designed and evaluated the thermal therapy delivered by both single- and multi-objective beamformers designed for various testbeds.
	The computational test beds were designed with varying fidelity to mimic breast tissue phantoms. Breast tissue can generally be classified as fatty or fibroglandular tissue, with fibroglandular tissue having a higher water content and therefore higher dielectric properties \cite{lazebnik2007largetissue}. 
	\subsection{Dielectric and Thermal Models}
	\label{sub_sec:fdtd_models}

	The dielectric properties of the various computational testbeds were represented by a single-pole Debye model, which provided an accurate representation of the dielectric behavior over the microwave frequencies relevant to this analysis. 
	The single-pole Debye model calculates the complex permittivity
	\begin{equation}
		\label{eq:debye}
		\epsilon(\omega) = \epsilon_{\infty} + \frac{\Delta\epsilon}{1 + j\omega\tau} + \frac{\sigma_{\mathrm{s}}}{j\omega\epsilon_0}
	\end{equation}
	where the $\Delta\epsilon = \epsilon_{\mathrm{s}} - \epsilon_{\infty}$, in which $\epsilon_{\infty}$ is the permittivity at the high frequency limit and $\epsilon_{\mathrm{s}}$ is the static permittivity, $\epsilon_0$ is the free-space permittivity, $\omega$ is the angular frequency, $\sigma_{\mathrm{s}}$ is the static conductivity and $\tau$ is the relaxation time. The effective conductivity, $\sigma_{\mathrm{eff}}$, is proportional to the imaginary part of $\epsilon(\omega)$, $\epsilon''$, through
	\begin{equation}\label{sigma}
		\sigma_{\mathrm{eff}} = \epsilon''\omega\epsilon_0
	\end{equation}
	The Debye parameters for fatty and fibroglandular breast tissue are taken from \cite{converse2006computational} and shown in Table \ref{tab:debey_param}.
	
	\begin{table}[tb]
		\ifcsname hlon\endcsname%
		\color{blue}%
		\fi%
		\caption{Single-Pole Debye Parameters of Breast Tissues at \SI{2.5}{\giga\hertz}~\cite{converse2006computational}}
		\label{tab:debey_param}
		\begin{center}
			\begin{tabularx}{\columnwidth}{p{0.25\linewidth}YY}
				\toprule[1pt]
				\midrule
				Parameter & Value in Fat & Value in Fibroglandular\\
				\midrule
				$\epsilon_{\infty}$ &  3.39 & 17.5\\
				$\Delta\epsilon$ &  2 & 31.6\\
				$\sigma_{\mathrm{s}}$ &  0.05 & 0.72\\			
				$\tau$ (ps) & 0.15 & 0.15\\			
				\midrule
				\bottomrule[1pt]	
			\end{tabularx}
		\end{center}
	\end{table}
	
	The hyperthermia performance of the near-field beamformer is characterized by first determining the beamformer-specific heating potential and then simulating the resulting steady-state temperature distribution within the media. The spatially and temporally changing temperature of the computational testbeds were represented using the Pennes bio-heat equation \cite{pennes1948analysis} similar to \cite{bernardi2003specific, zastrow20103d,2011_Zastrow_time_multiplexing,burfeindt2011microwave}. It describes the temporal evolution of heat within tissue depending on the capability of the tissue to store thermal energy. The steady state temperature map $T(\mathbf{r})$ is generated in the thermal model by 
	\begin{multline}
		\label{eq:pennes_bio_heat}
		C_p(\mathbf{r})\rho(\mathbf{r}) \frac{\partial T(\mathbf{r})}{\partial t} = \nabla \cdot \left( K(\mathbf{r}) \nabla T(\mathbf{r}) \right) + A_0(\mathbf{r}) + \\
		Q(\mathbf{r}) - B(\mathbf{r})(T(\mathbf{r}) - T_B) \quad (\mathrm{Wm^{-3}})
	\end{multline}
	where $C_p$ is the specific heat, $\rho$ is the density of the tissue, $K$ is the thermal conductivity, $A_0$ and $B$ are the metabolic heat generation and the capillary blood perfusion coefficients, respectively, $T_B$ is the blood temperature with the normal body temperature of \SI{37}{\degreeCelsius}, and the electromagnetic heating potential $Q$ \SI{}{\watt\per\meter\squared}  is calculated from a preliminary FDTD simulation that evaluates beamformer power deposition and is given by 
	\begin{equation}
		\label{eq:heat_p}
		Q_{i, j} = \frac{1}{T} \sum_{n = n_{m} - n_{p}}^{n_{m}} \left( {\mathbf{E}_{i, j}^{n}} \cdot {\mathbf{J}_{i, j}^{n}} \right) \Delta t
	\end{equation}
	where \textbf{E} is the electric field intensity, \textbf{J} is the electric current density, $i$ and $j$ are the computational grid indices for the 2-D test environment, $T$ is the period of the transmitted signal, $n_{p}$ is the number of steps in one period, and $n_{m}$ is the maximum simulation time step. $\Delta t$ is the {time step used to discretize the FDTD solver}. Since $\vec{J} = \sigma\vec{E}$, \eqref{eq:heat_p} can be expressed as
	\begin{equation}
		\label{eq:heat_p_v2}
		\begin{split}
			Q_{i, j} &  = \frac{1}{T} \sum_{n = n_{m} - n_{p}}^{n_{m}} \left( {\mathbf{E}_{i, j}^{n}} \cdot \sigma_{i, j} {\mathbf{E}_{i, j}^{n}} \right) \Delta t \\
			& = \frac{1}{T} \sum_{n = n_{m} - n_{p}}^{n_{m}} \sigma_{i, j}\left( \mathbf{E}_{i, j}^{n} \right)^{2}\Delta t
		\end{split}
	\end{equation}
	where $\sigma = \sigma_{\mathrm{eff}}$ is calculated from \eqref{sigma} for each cell in the grid. The heating potential is then scaled to the required input power needed to achieve the desired temperatures, representing a uniform scale on antenna power during the hyperthermia treatment. The relevant thermal parameters are detailed in Table \ref{tab:thermal_param} and are taken from \cite{converse2006computational} and \cite{gosalia2004thermal}. The convective heat boundary coefficient between the tissue and the surrounding media in the simulations is given by $H$ and has a value of \SI{5}{\watt \per\meter\squared \per\degreeCelsius} for a tissue-air \cite{seagrave1971biomedical} and \SI{300}{\watt \per\meter\squared \per\degreeCelsius} for a tissue-water interface \cite{converse2006computational} respectively.
	
	\begin{table}[tb]
		\ifcsname hlon\endcsname%
		\color{blue}%
		\fi%
		\caption{{Thermal Properties}}
		\label{tab:thermal_param}
		\begin{center}
			\begin{tabularx}{\columnwidth}{p{0.26\linewidth}p{0.15\linewidth}YY}
				\toprule[1pt]
				\midrule
				Parameter & Fat~\cite{converse2006computational} & Fibroglandular~\cite{gosalia2004thermal} & Water~\cite{converse2006computational} \\
				\midrule
				$C_p$ (\SI{}{\joule \per\kilogram \per\degreeCelsius}) & 2279 & 3600 & 4186\\[0.5ex]
				$K$ (\SI{}{\watt \per\meter \degreeCelsius}) & 0.306 & 0.5 & 0.6\\ [0.5ex]
				$\rho$ (\SI{}{\kilogram \per\meter}) & 1069 & 1050 & 1000\\[0.5ex]
				$A_0$ (\SI{}{\watt \per\meter\cubed}) & 350 & 690 & --\\[0.5ex]
				$B$ (\SI{}{\watt \per\meter\cubed \per\degreeCelsius}) & 2229 & 2700 & --\\
				\midrule
				\bottomrule[1pt]	
			\end{tabularx}
		\end{center}
	\end{table}
			
	

	\subsection{Computational Testbeds}
	\label{sub_sec:testbeds}
	
	We evaluated our thermal therapy performance in computational testbeds that represented homogeneous and heterogeneous media. Additionally, we evaluated the performance of the beamformer in both simple geometries and anatomically realistic phantoms. All computational testbeds were used in two-dimensional FDTD simulations and therefore represent an infinitely long cylinder with the given cross section.
	
\begin{figure}[t!]
		\centering
		\includegraphics[width=1\linewidth]{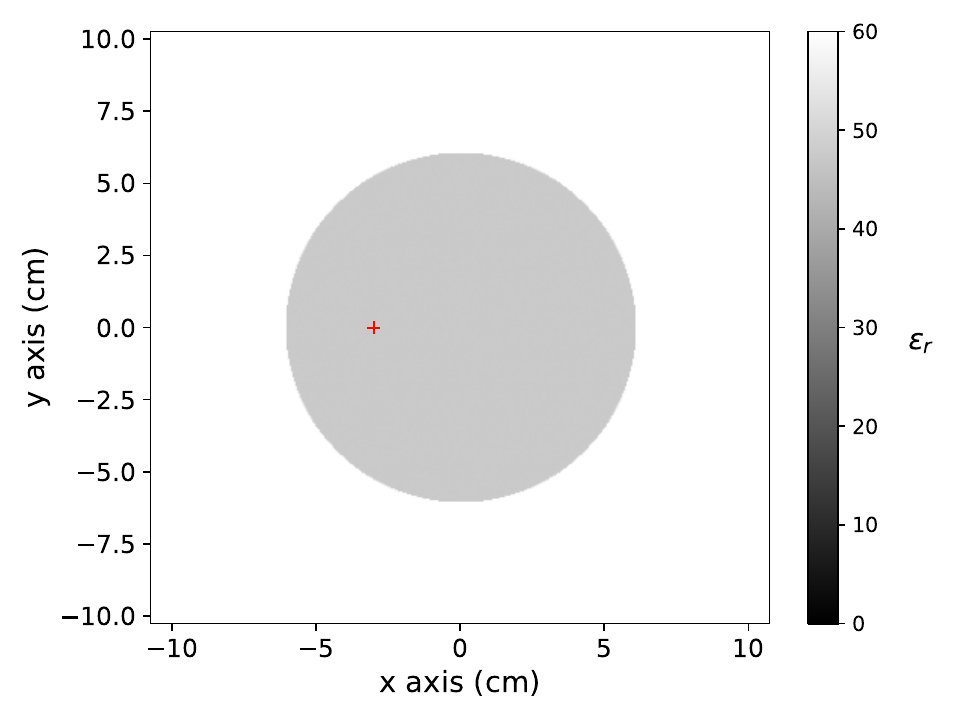}
		\caption{Homogeneous fibroglangular tissue media setup with no dielectric discrepancy. The red $+$ indicates the beamforming target location.}
		\label{fig:so_hom_no_ablation} 
		\vspace{-\baselineskip}
	\end{figure}

	The first computational testbed we used to evaluate our beamformer performance was a dielectrically and thermally homogeneous testbed with a simple cylindrical geometry surrounded by air. The dielectric properties for the homogeneous phantoms were selected to mimic either fatty or fibroglandular breast tissue with dielectric properties mentioned above. The homogeneous phantom was designed to have a radius of \SI{6}{\centi \meter}. The simple computational testbed used a uniform spatial grid of 400 by 400 cubic Yee cells with dimensions of $\delta x = \delta y = \SI{0.5}{\milli \meter}$. {This provides a grid resolution relating to a sampling density of $N_{\lambda}$ $\sim$ \SI{28}{} in fibroglandular breast tissue at \SI{2.5}{\giga \hertz}.} An example of the simple homogeneous fibroglandular tissue test environment is shown in Fig. \ref{fig:so_hom_no_ablation}. Fatty breast tissue has a much lower relative permittivity than fibroglandular breast tissue due to its much lower water content and therefore represents a much finer grid resolution. The computational grid is terminated by uniaxial perfectly matched layer (UPML) absorbing boundary conditions.  To ensure stability the Courant factor is chosen as $S = \SI{0.7}{}$.
	
\begin{figure}[t!]
		\centering
		\includegraphics[width=1\linewidth]{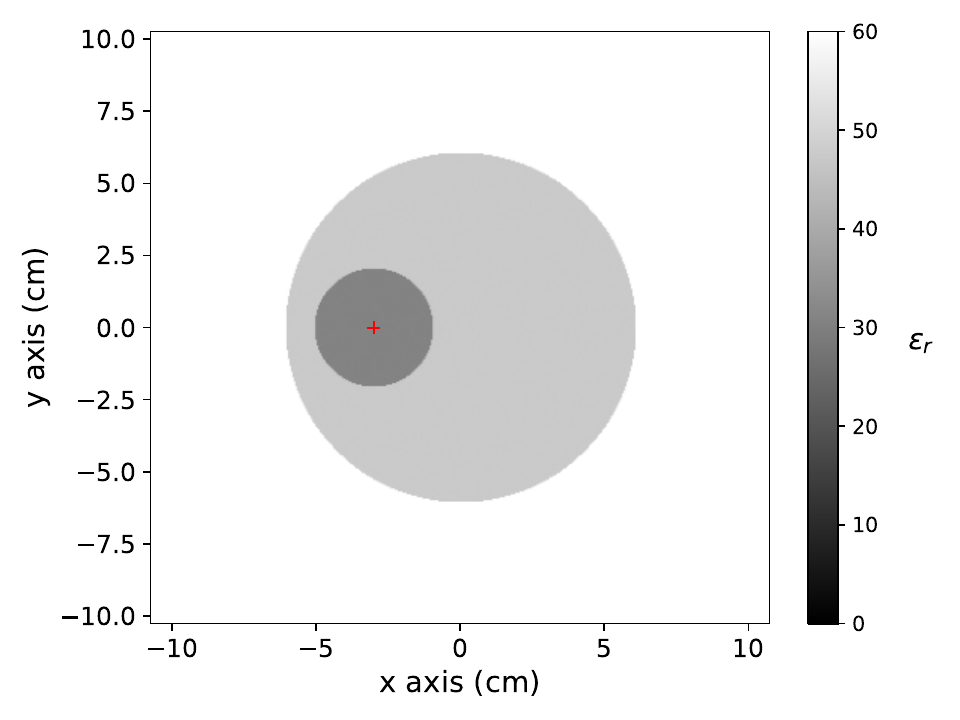}
		\caption{Inhomogeneous media setup with Debye dielectric parameters decreased up to \SI{65}{\percent} at the treatment region of \SI{2}{\centi \meter} radius around the focus location. The red $+$ indicates the beamforming target location.}
		\label{fig:so_inhom_only_ablation} 
		\vspace{-\baselineskip}
	\end{figure}

\begin{figure}[t!]
		\centering
		\includegraphics[width=1\linewidth]{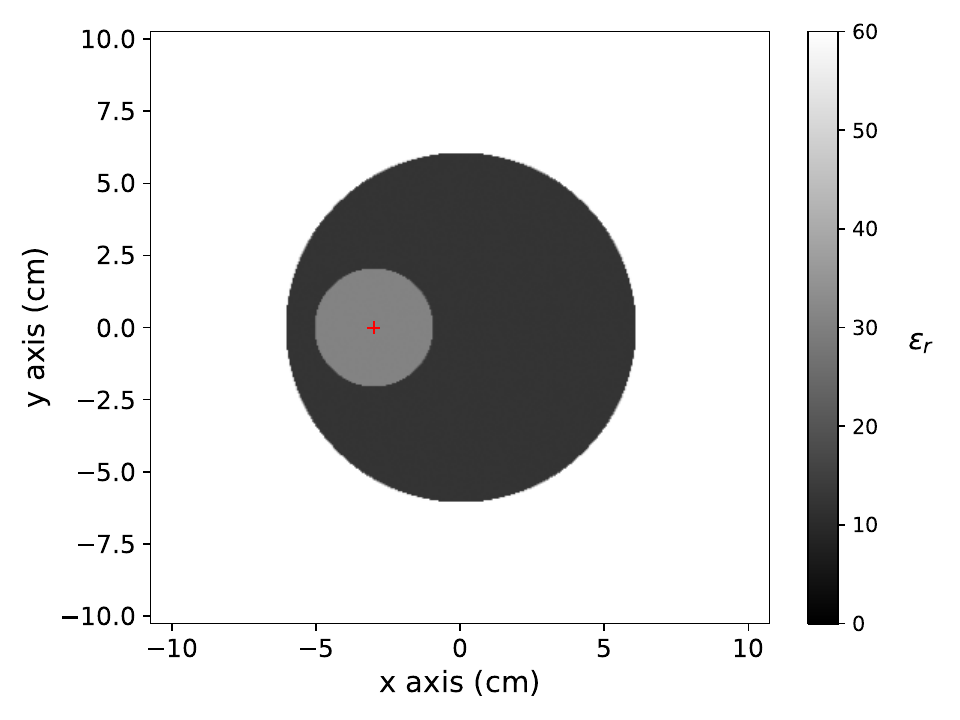}
		\caption{Inhomogeneous media setup with Debye dielectric parameters decreased up to 65\% at the treatment region of \SI{2}{\centi \meter} radius around the focus location and the surrounding decreased to 26\%. The red $+$ indicates the beamforming target location.}
		\label{fig:so_inhom_surround_ablation} 
		\vspace{-\baselineskip}
	\end{figure}

\begin{figure}[t!]
		\centering
		\includegraphics[width=1\linewidth]{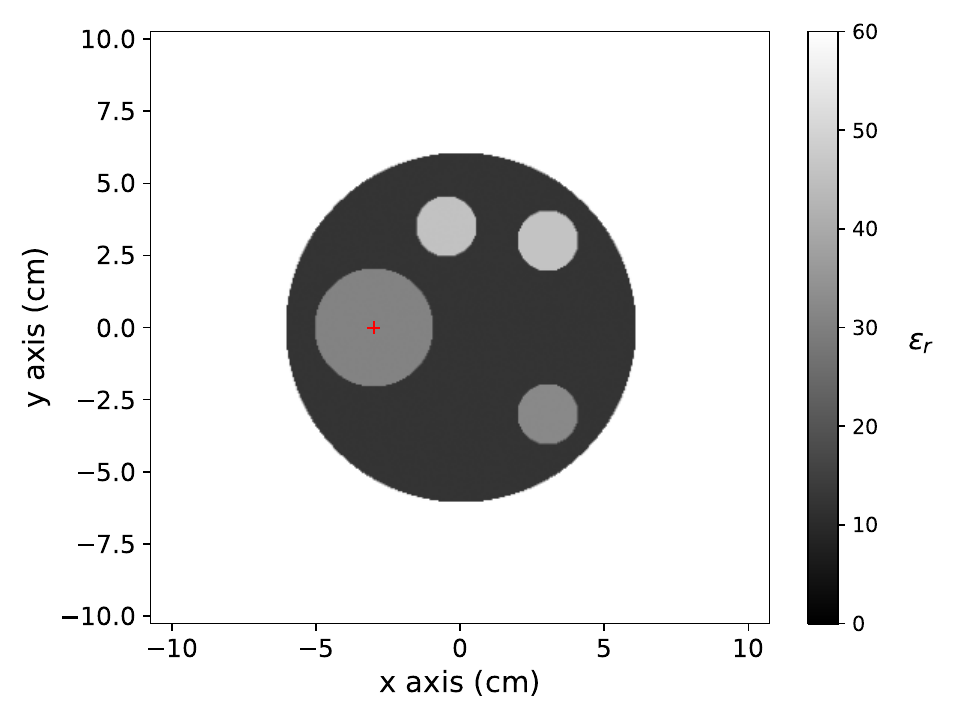}
		\caption{Inhomogeneous media setup with Debye dielectric parameters decreased up to 65\% at the treatment region of \SI{2}{\centi \meter} radius around the focus location and the surrounding decreased to 26\%. Three other hotspots are depicted with a decreasing rate of 1\%, 3\% and 4\% for hotzone A, hotzone B and hotzone C. The red $+$ indicates the beamforming target location.}
		\label{fig:so_inhom_surround_ablation_hz} 
		\vspace{-\baselineskip}
	\end{figure}

	A degree of inhomogeneity was next introduced in the simple testbeds in order to begin mimicking scenarios where the dielectric properties change in response to heating, which will disrupt the beamforming, the weights for which are based on the initial, unheated dielectric properties. At \SI{2.5}{\giga \hertz} the dielectric properties decrease with increased temperature~\cite{haldes2025temperature}, thus the parameters were incrementally decreased non-uniformly in distinct regions of the initial homogeneous test environment; note that the variations in the Debye parameters do not necessarily correspond to a proportional change in the dielectric properties. The dielectric properties for each iteration of the FDTD algorithm for the incrementally modified Debye parameters were calculated by \eqref{eq:debye}. The near-field beamformer was evaluated in three different simple, inhomogeneous scenarios:
	\begin{itemize}
		\item[a.] We decrease the Debye parameters of only the radius treatment region by \SI{5}{\percent} to \SI{35}{\percent}. Further reduction would result in decreasing the properties below zero which is not physically realistic. Fig. \ref{fig:so_inhom_only_ablation} shows the permittivity map of the test media when the treatment region has been decreased to \SI{35}{\percent} while the surrounding media remains unchanged.
		\item[b.] We decrease both the treatment region and the surrounding region by different percentages. The treatment region is decreased by \SI{5}{\percent} as before up to \SI{35}{\percent} while the surrounding is decreased by \SI{2}{\percent} up to \SI{74}{\percent}. Fig. \ref{fig:so_inhom_surround_ablation} shows the permittivity map when the Debye parameters of the treatment region is at \SI{35}{\percent} and the surrounding region is at \SI{74}{\percent} of the initial map. 
		\item[c.] We selected three additional arbitrarily located regions with radii of \SI{1}{\centi\meter} and decreased their Debye parameters by \SI{1}{\percent}, \SI{3}{\percent} and \SI{4}{\percent}. Fig. \ref{fig:so_inhom_surround_ablation_hz} shows this permittivity map with the target and the three additional dielectric inclusions.
	\end{itemize} 
	
	One of the challenges of a non-invasive system design is that it is not always possible to report the changes of the dielectric properties in real time. Using spatially averaged properties, the authors in \cite{zastrow20103d} designed a beamformer and demonstrated that microwave focusing can be maintained even in highly heterogeneous tissue environments. We employ a similar method here to design our partial knowledge enhanced beamformer. We use the heterogeneous environment shown in Fig. \ref{fig:so_inhom_surround_ablation_hz} and take the spatial average of the dielectric properties to construct the propagation model and obtain the beamformer weights at each iteration of gradually changing dielectric properties.
	
\begin{figure}[t!]
		\centering
		\includegraphics[width=1\linewidth]{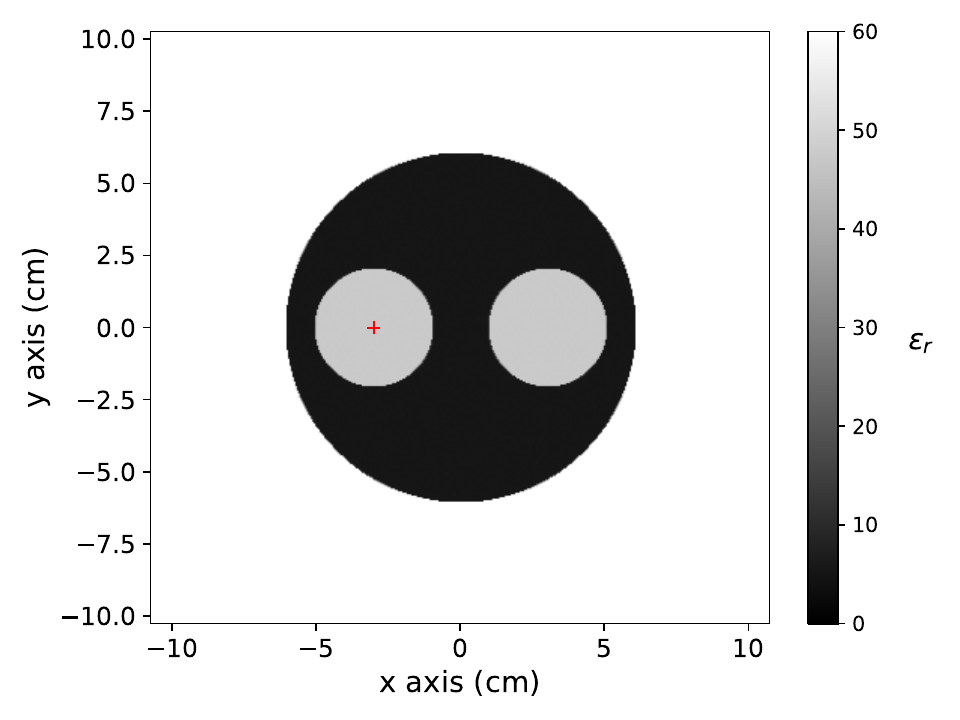}
		\caption{Heterogeneous media setup with Debye dielectric parameters similar to that of a class two MRI-derived breast phantom. The red $+$ indicates the beamforming target location.}
		\label{fig:inhom_mo} 
		\vspace{-\baselineskip}
	\end{figure}

	A second testbed was based on two fibroglandular tissue regions embedded in fatty tissue. The test environment is shown in Fig. \ref{fig:inhom_mo}. The heterogeneous media consisted of \SI{12}{\centi\meter} diameter of fatty tissue with two \SI{2}{\centi\meter} radius inclusions of fibroglandular tissue within it at $\left(-\SI{3}{\centi\meter}, 0\right) $ and $\left(\SI{3}{\centi\meter}, 0\right) $, respectively. We presume that the tumor is located on the left, at $\left(-\SI{3}{\centi\meter}, 0\right) $ making it the primary focus location while the fibroglandular tissue on the right is a secondary beamformer target location where nulls will be steered if unwanted heating occurs.  
	
\begin{figure}[t!]
		\centering
		\includegraphics[width=1\linewidth]{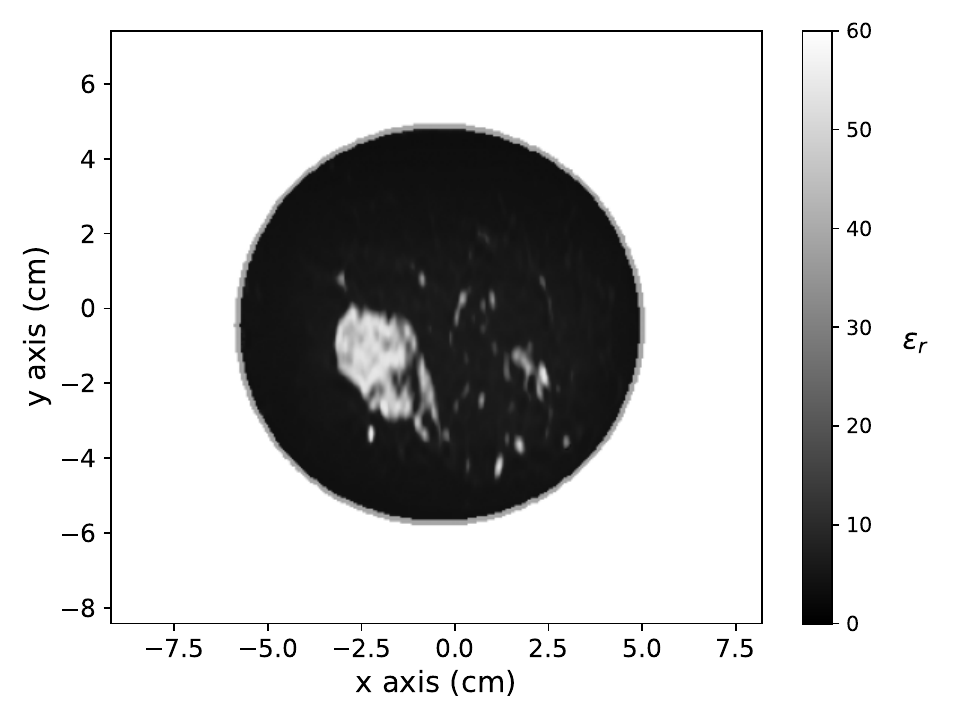}
		\caption{MRI derived class II breast phantom with scattered fibroglandular tissue. The immersion media is water.}
		\label{fig:virtual_patient}
		\vspace{-\baselineskip}
	\end{figure}

	Lastly, we evaluated an anatomically, dielectrically, and thermally realistic computational testbed, hereafter referred to as the "realistic phantom". Realistic breast tissue phantoms were derived by the authors in \cite{zastrow2008development} from MRIs taken of volunteers' healthy breast tissue. The phantoms represent a wide range of variability in patient-to-patient breast tissue density. In this study, the selected phantom had a breast tissue density that was classified as scattered fibroglandular. This selected model represents a patient that distinctly has both fibroglandular and fatty tissue regions, as opposed to patients with more densely fatty or fibroglandular breast tissue. This computational testbed was comprised of the MRI-derived anatomically realistic phantom immersed in dielectrically and thermally accurate water, it is shown in Fig. \ref{fig:virtual_patient}.
	
	\subsection{Beamforming}
	\label{sub_sec:bf_setup}
	
	The beamformers are designed using a circular antenna array around the tissue region with a \SI{7}{\centi \meter} radius for the simple models and with a \SI{6.5}{\centi \meter} radius for the realistic phantom. It was assumed that array elements were electric current sources operating at a carrier frequency of \SI{2.5}{\giga \hertz}. Time domain analysis was used for both acquiring the beamformer weights and their performance analysis given by the resulting simulated power deposition data. Beamforming weights were obtained using a time-reversal process similar to that used in~\cite{mukerjee2022microwave}. A pulsed waveform with a bandwidth of 750 MHz (30\% fractional bandwidth) was emitted from the desired focus location in the media, which is received at all receiver locations surrounding the media, at which the complex channel weights are determined based on the propagation of the pulse. 
	The wave propagation is simulated using the FDTD solver for the $TM_z$ mode. The signals at each receiving antenna location are converted to frequency domain using discrete Fourier transformation (DFT) and normalized to a reference antenna, which was the antenna to the right of the tissue region at $(x = \SI{7}{\centi\meter}, y = \SI{0}{\centi\meter})$. The frequency domain signals provide the channel weights, which are then used to generate the CW signals at \SI{2.5}{\giga\hertz}, which, with the weights determined from the DFT, are propagated back through the media and converge at the focus region. The FDTD algorithm was used to track the electromagnetic field propagation and evaluate the power deposition in the entire tissue region using \eqref{eq:heat_p_v2}. 
	
	Both single objective and multi-objective beamformers were designed. For the single objective beamformer the phases of the beamformer weights are a complex conjugate match of the channel weights. 
	The multi-objective beamformer is designed using a modified version of the traditional linearly constrained minimum power (LCMP) far-field multi-objective beamforming algorithm~\cite[p.~513--553]{trees2002optimum}. The feasibility of using the LCMP algorithm for free-space near-field beamsteering application has already been experimentally demonstrated in~\cite{bhattacharyya2023multiobjective, bhattacharyya2024multiobjective} for a distributed phased array based on wireless frequency syntonization and time synchronization, respectively.
	
	\label{sub_sub_sec:lcmp_eq}
	The conventional LCMP method employs a null constraint matrix $\mathbf{C}$ to compute the requisite weights for multi-objective beamforming. The $N \times M$ constraint matrix $\mathbf{C}$, where $N$ is the number of transmitters and $M$ the number of objectives (focuses and/or nulls), ensures linear independence among its columns, provided the condition $M = N - 1$ is maintained. The beamforming and null steering weights are then determined by
	\begin{equation}
		\label{eq:w_lcmp}
		\mathbf{w}_\mathrm{lcmp} = \mathbf{g}^H(\textbf{C}^H\textbf{C})^{-1}\textbf{C}^H
	\end{equation}
	where $(\cdot)^H$ denotes the Hermitian transpose. The constraint matrix $\mathbf{C}_{N \times M}$ comprises vectors containing the channel weights evaluated using the FDTD algorithm for each individual objective, that is, focus or null. The vector $\mathbf{g}~=~[B_1\,|\,...\,|\, B_m \,|\,...\,|\, B_M]^T$ is used to define the focii $\left( B_m=1 \right) $ and nulls $\left( B_m=0 \right) $. 
	
	The LCMP beamformer is a distortionless process that maintains power at the main focus while being able to steer nulls arbitrarily without requiring any feedback from the test environment. All it requires is the pre-determined location of the focus and nulls and the channel dielectric characteristics acquired by the FDTD simulations.
	For simplicity of the system we assume that all the array elements transmit with the same unit amplitude. The estimated weights, or phases, are used to design the CW signals emitted from the elements to achieve focusing and nulling. The spatial distribution of the power deposition or the heating potential from the resulting electric fields at each cell in the grid is used to evaluate the beamformer performance.
	
	\section{Single Objective Beamformer}
	\label{sec:so_results}
	\subsection{Homogeneous Media}
	\label{sub_sec:so_hom}
	
\begin{figure*}
		\centering
		\includegraphics[width=1\linewidth]{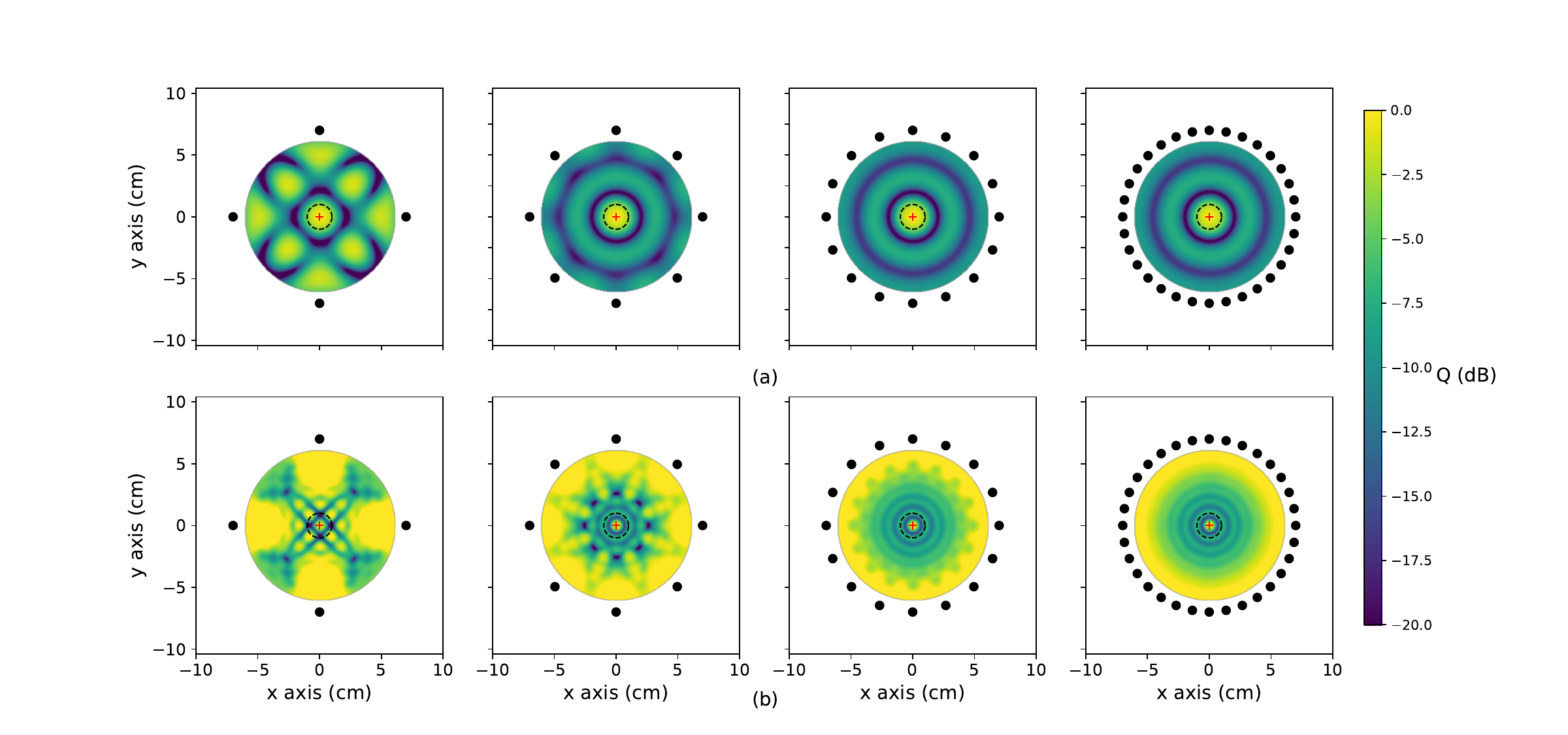}
		\caption{Single objective power deposition data for (a) fat tissue and (b) fibrograndular tissue in air with focus cell location at the center for four, eight, sixteen and thirty-two antenna elements normalized by the power at the focus location (0, 0).}
		\label{fig:so_pd_fat_fibro_center} 
		\vspace{-\baselineskip}
	\end{figure*}

\begin{figure*}[t!]
		\centering
		\includegraphics[width=1\linewidth]{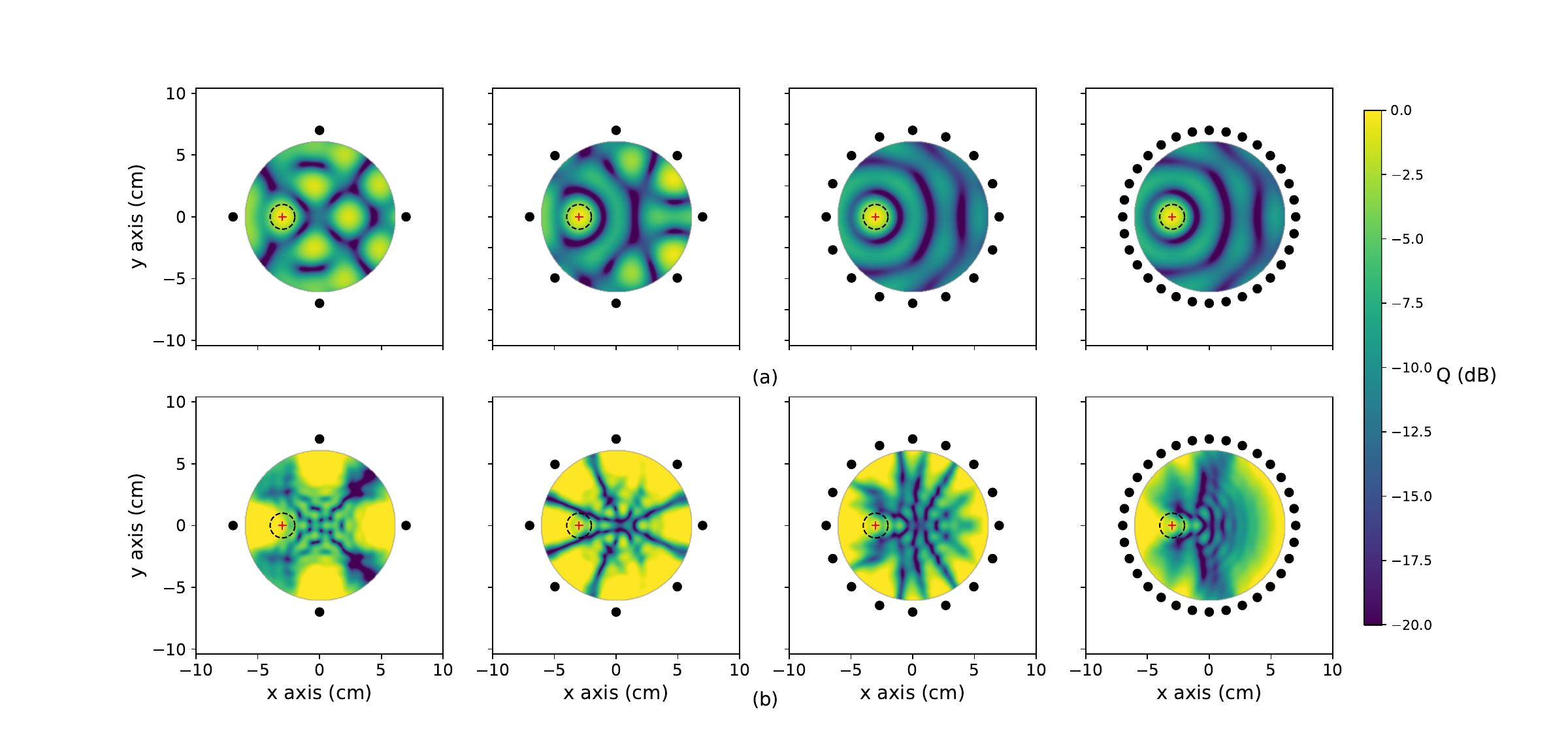}
		\caption{Single objective power deposition data for (a) fat tissue and (b) fibrograndular tissue with focus cell location off-center at  (x = \SI{-3}{\centi\meter}, y = 0) for four, eight, sixteen and thirty-two antenna elements in air  normalized by the power at the focus location.}
		\label{fig:so_pd_fat_fibro_off_center} 
	\end{figure*}

\begin{figure*}[t!]
		\centering
		\includegraphics[width=1\linewidth]{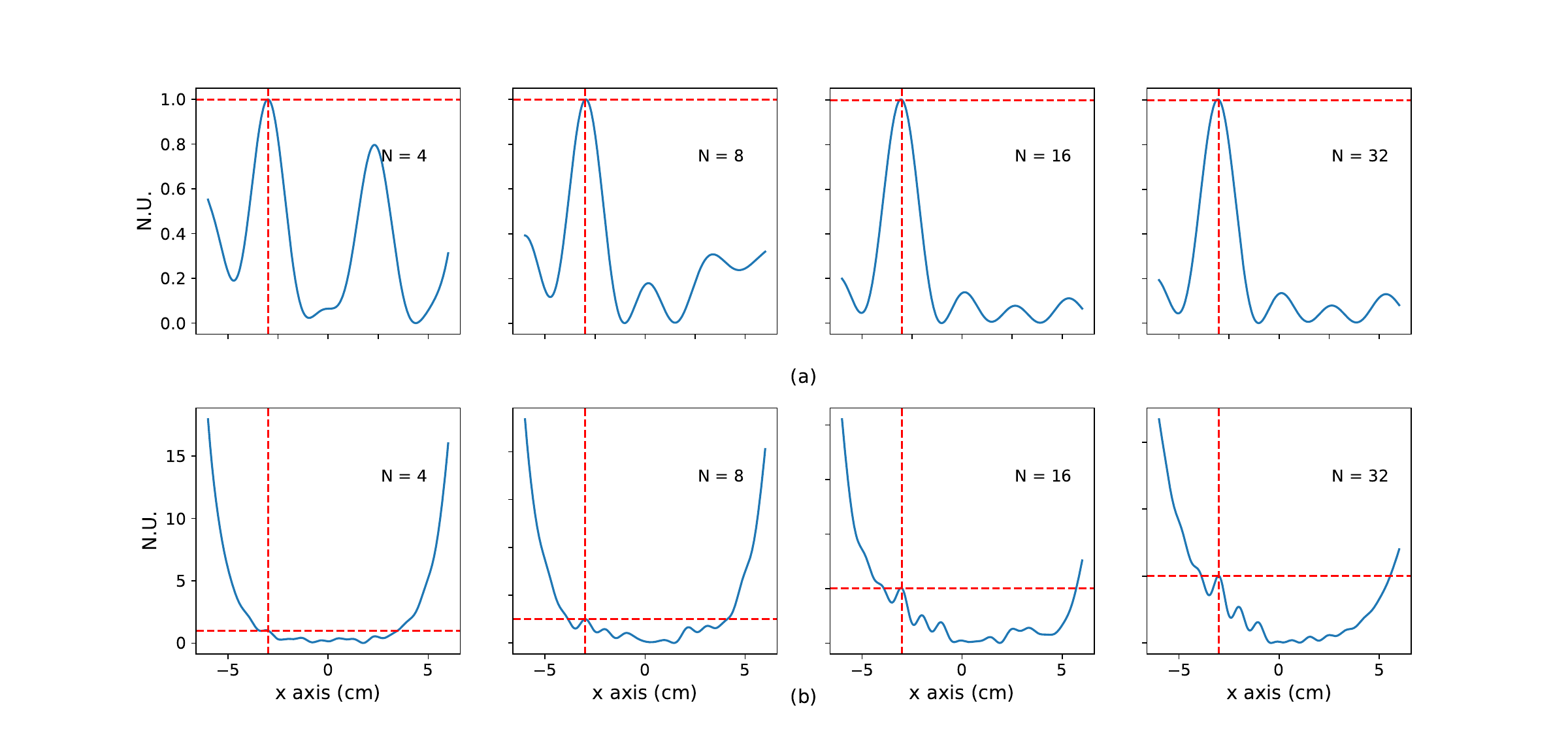}
		\caption{Power deposition profile along the x-axis at y = \SI{0}{} when the focus is off-center at (x = \SI{-3}{\centi\meter}, y = 0) for both (a) fat and (b) fibrograndular tissue with number of transmitters specified by $\mathrm{N}$.}
		\label{fig:so_yslice_fat_fibro_off_center} 
	\end{figure*}

	A simple case of homogeneous fat or fibroglandular tissue was first evaluated to determine the effects of the number of antennas. Since tumor cells are usually found in the fibroglandular tissue, our focus is mostly on fibroglandular tissue models. The setup is given by Fig.~\ref{fig:so_hom_no_ablation}. 
	We varied the number of antenna elements in the array and examined the capability of focusing at the center (0, 0) by studying the resulting power deposition images from the FDTD simulations as shown in Fig.~\ref{fig:so_pd_fat_fibro_center} for both fat and fibroglandular tissue surrounded by air. 
	It is seen that for both fat and fibroglandular tissue there is considerable power absorbed outside the treatment region for the four and eight array elements cases. The best focusing is achieved by the thirty-two element array, but, for a tissue radius of \SI{6}{\centi \meter} it is not physically feasible to place antenna elements that close together without using small elements that incur significant mutual coupling. We determined that the best trade-off between practicality and maintaining a narrow focus is by using 16 elements.          
	\begin{figure*}[t!]
		\centering
		\includegraphics[width=1\linewidth]{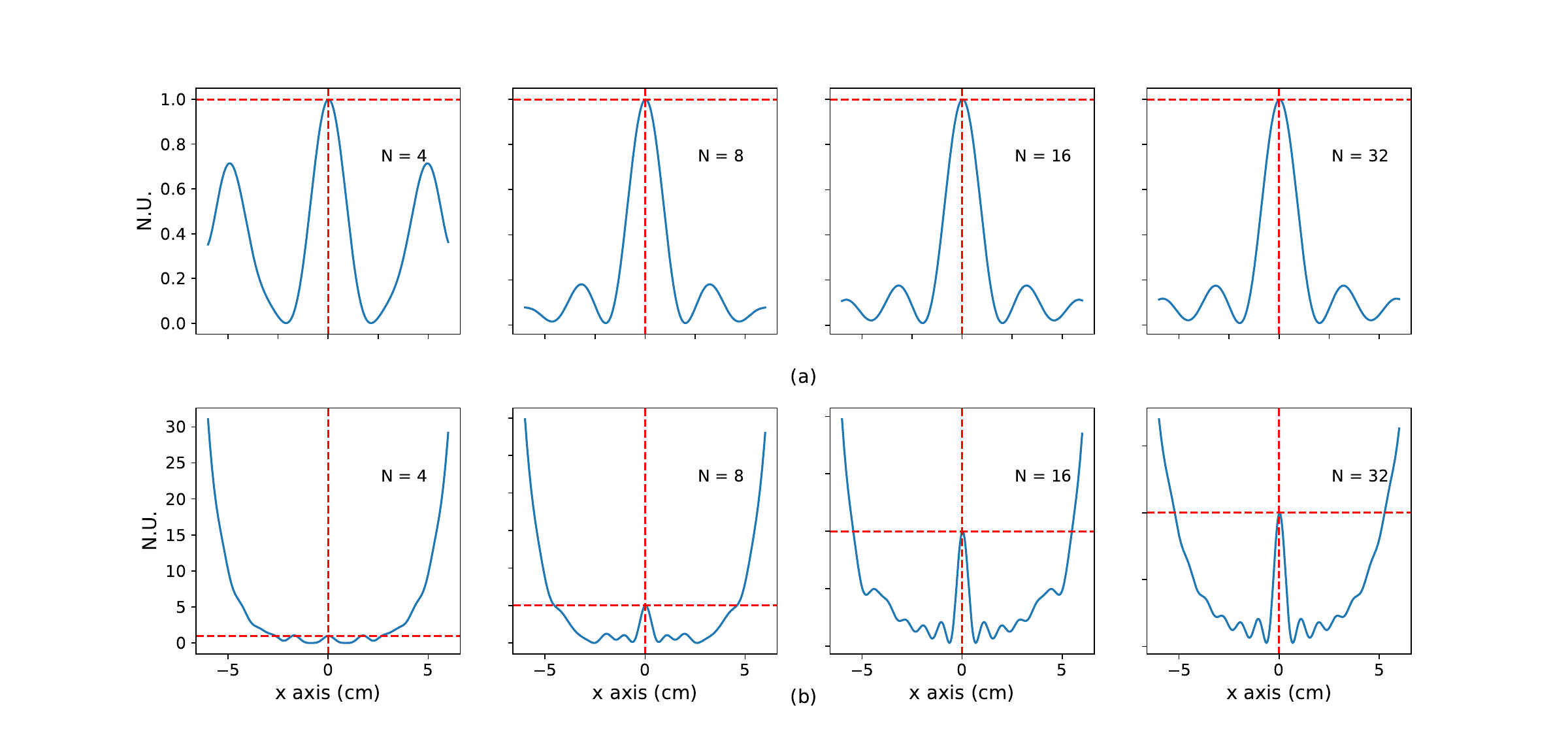}
		\caption{Power deposition profile along the x-axis at y = \SI{0}{} when the focus is at the center for (a) fat tissue and (b) fibroglandular tissue for $\mathrm{N}$ number of transmitters.}
		\label{fig:so_yslice_fat_fibro_center} 
	\end{figure*}
	We then moved the focusing location off-center to (\SI{-3}{\centi \meter}, 0) and find that the previous observations hold true (Fig.~\ref{fig:so_pd_fat_fibro_off_center}). Another interesting inference that we can draw from this set of tests is that, when we look at the 1-dimensional (1D) slices of the power deposition along the x-axis at $y = \SI{0}{}$, as in Figs.~\ref{fig:so_yslice_fat_fibro_center} and \ref{fig:so_yslice_fat_fibro_off_center}, is that, for fatty tissue the sidelobe levels or the absorbed power outside the treatment region is lower by approximately \SI{80}{\percent} than in the  treatment region, especially for the sixteen element setup. This is evident for both the center and off-center focusing cases. We also see that the power at the edge is low for fatty tissue, whereas, for fibroglandular tissue it is high. 
	
	We can also show that the efficiency of the beamformer is highly dependent on the impedance matching between the incident wave and the tissue boundary. The impedance of most types of tissue is usually lower than that of air. Due to higher water content fibroglandular tissue is more lossy than fatty tissue.
	This impedance mismatch for air-tissue boundary results in unwanted power being deposited at the tissue edge. 
	The substantial impedance mismatch between air and fibroglandular tissue, resulting from an air gap between the beamformer elements and the body, leads to significant reflection of the incident electric field at the tissue boundary, limiting its penetration into the tissue region and causing unwanted energy deposition at the edge. 
	
\begin{figure}
		\centering
		\includegraphics[width=1\linewidth]{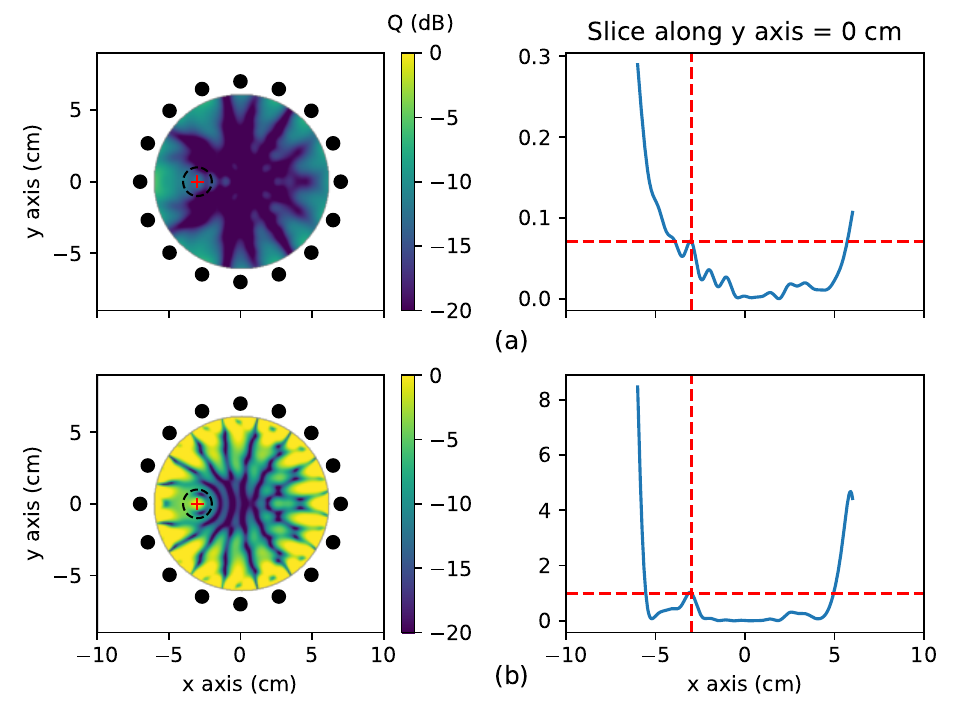}
		\caption{Power deposition in tissue and the 1D slice along the x-axis at y = 0 of the target to show the difference in the amount of power absorbed by the tissue when the surrounding media is: (a) air and (b) water. The data is normalized by power deposited at the focus when the surrounding media is water. The target region of \SI{1}{\centi \meter} radius around the focus is denoted by the dashed line.}
		\label{fig:sar_surround_media_comparison} 
	\end{figure}

	We studied the effects of air and water as surrounding media more closely for the 16 element antenna array beamformer with homogeneous fibroglandular media. The effects of both on the power deposition can be observed in Fig.~\ref{fig:sar_surround_media_comparison}. The power deposiiton for both the cases were normalized using the power deposited at the focus cell when the surrounding media is water for a fair comparison between the two. The slices taken along the x-axis at $y = 0$ show that the power deposition at the focus is an order of magnitude lower for media cylinder in air than in water as a result of better impedance matching at the tissue. The total deposited power is also lower by the same amount. 
	
\begin{figure}
		\centering
		\includegraphics[width=1\linewidth]{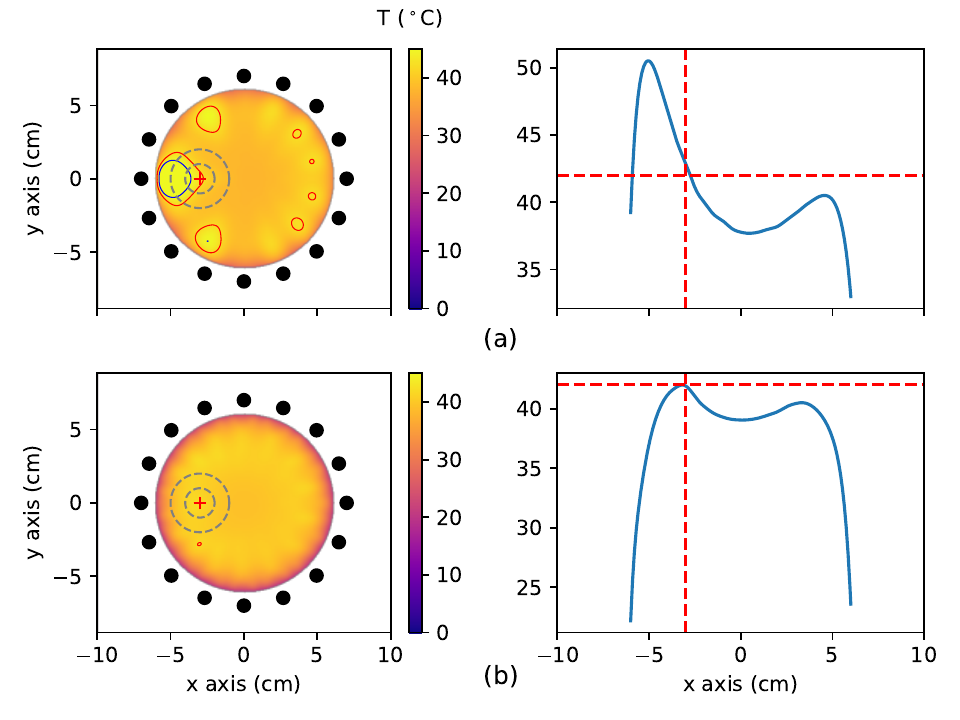}
		\caption{Thermal map and the 1D slice along the x-axis at the y = 0 of the target to show the comparison by the tissue when the surrounding media is: (a) air and (b) water. The data is normalized by power deposited at the target when the surrounding media is water. The target region of \SI{1}{\centi \meter} and the treatment region of \SI{2}{\centi \meter} radius around the focus} is denoted by the dashed line.
		\label{fig:tm_surround_media_comparison} 
	\end{figure}

	An additional benefit of using water as the surrounding media is that it can be used to actively cool the temperatures at the boundary of the tissue by maintaining a constant external water temperature. In our models we used \SI{15}{\degreeCelsius}, in order to mitigate the unwanted heating resulting from the energy absorption. Similar to Fig.~\ref{fig:sar_surround_media_comparison}, Fig.~\ref{fig:tm_surround_media_comparison} shows the temperature distribution of the fibroglandular tissue when it has air and water as the surrounding media. We take the power deposition data from Fig.~\ref{fig:sar_surround_media_comparison} for the two cases and then scale them by the same factor to obtain the input heating potential for the FDTD-thermal solver. It is evident that for the same scaling factor the media surrounded by air shows much higher edge heating, $\sim$ \SI{10}{\degreeCelsius} more than the media surrounded by water. This is a result of both the electromagnetic coupling at the media boundary and the active cooling effects of water.
	\begin{figure}
		\centering
		\includegraphics[width=1\linewidth]{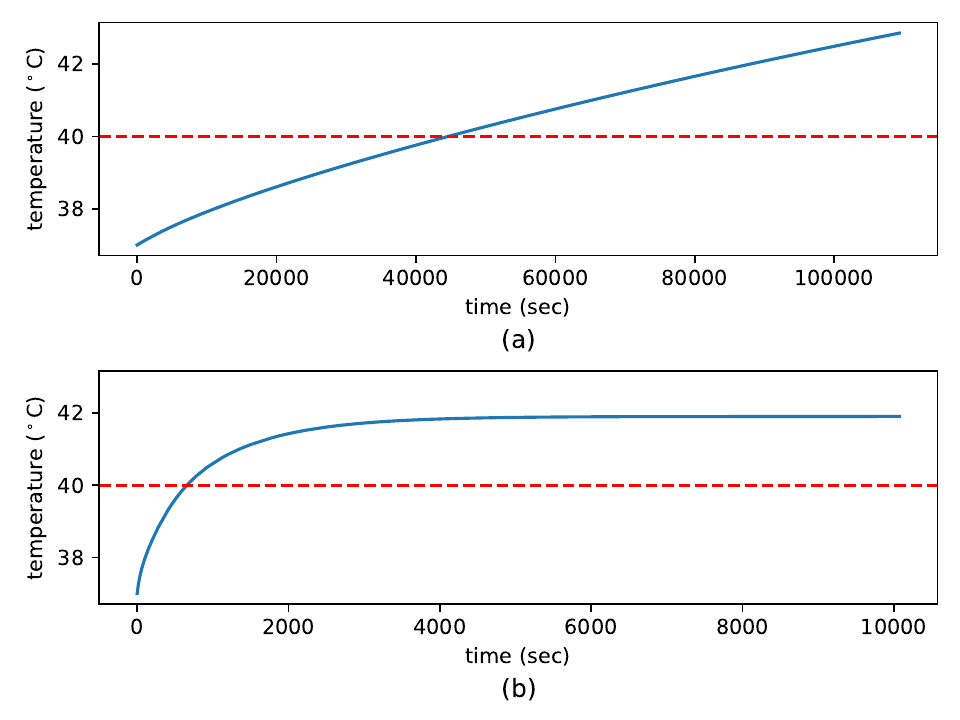}
		\caption{Time taken for the focus cell to reach ablation temperatures when the surrounding media is: (a) air and (b) water. The red line is at \SI{43}{\degreeCelsius} depicts the time at which the cell ablation temperature is reached.}
		\label{fig:time_temp_surround_media_comparison} 
	\end{figure}
	In Fig.~\ref{fig:time_temp_surround_media_comparison} the thermal nature of the media over time is analyzed. We specifically look at the time taken to reach thermal steady-state at the focus for both the test cases, for the same scaling factor of heating potential. The tissue surrounded by water reaches steady-state by \SI{11.03}{\minute}, while for air as surrounding media it does not reach steady-state. Hence, to optimize power deposition to the focus and minimize edge heating through effective active cooling, we focus solely on water as the surrounding medium for our analysis for the rest of the paper.
	
	\subsection{Inhomogeneous Media}
	\label{sub_sec:so_inhom}
	The permittivity maps for the three inhomogenous test cases evaluated in this section are shown in Figs.~\ref{fig:so_inhom_only_ablation}--\ref{fig:so_inhom_surround_ablation_hz}. These test environments are used to analyze three types of beamformer design: the static beamformer, the ideal beamformer and the partial knowledge beamformer.   
	\begin{figure}[t!]
		\centering
		\includegraphics[width=1\linewidth]{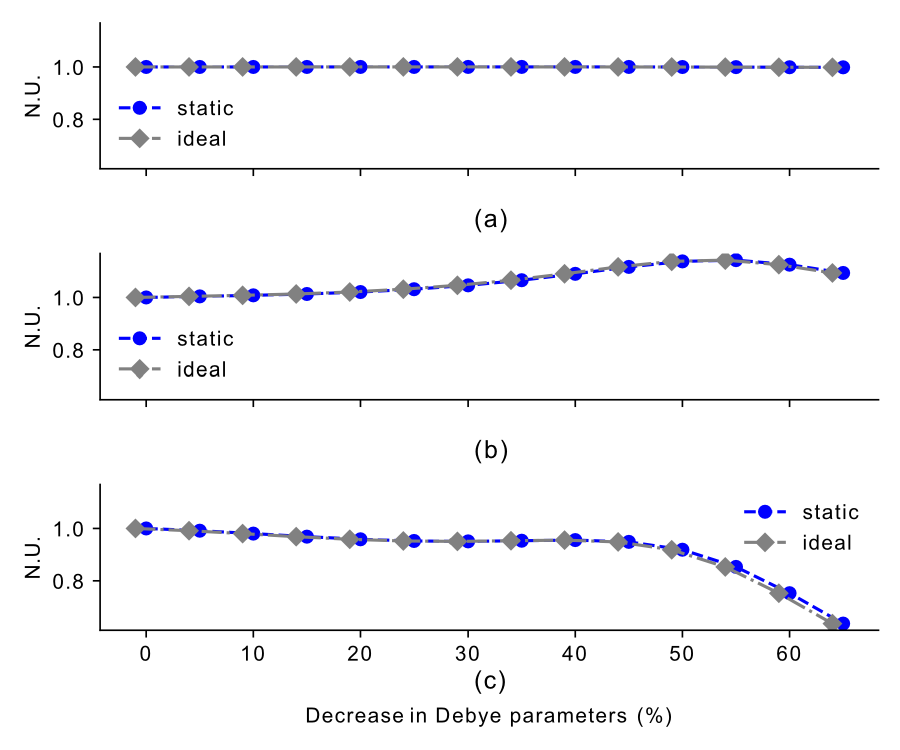}
		\caption{{Changes in the absorbed power by the (a) total media, (b) treatment region and (c) target cell for the ideal and static beamformer when only the Debye parameters of the treatment region is reduced by \SI{5}{\percent}. The power levels are normalized to the initial static case.}}
		\label{fig:pwr_analysis_only_ablation} 
	\end{figure}
	We start with a simple case of changing only the treatment region Debye properties by \SI{5}{\percent} up to \SI{35}{\percent}, to compare a static beamformer to an ideal beamformer that adjusts its weights appropriately to the changes in the dielectric properties. We analyze the performance of the beamformer by quantifying the power deposited in the total tissue cylinder, the power deposited in the treatment region, and the power at the target cell (the focal point).  The power is normalized by the baseline case which is the initial testbed with no changes made to the dielectric properties. This holds true for the all subsequent poswer deposition analysis.
	Fig.~\ref{fig:pwr_analysis_only_ablation} shows the change in the normalized power deposition as a function of the changing Debye parameters for the two beamformers. It can be seen that for this case, where the media changes occur in a small area of about a radius of $\sim$ \SI{2}{\centi\meter} surrounding the target cell, the static beamformer performs similar to the ideal beamformer. This indicates that if the only changes due to heating are in the target region then a static beamformer will likely perform well as well as the ideal beamformer. Since the power deposition is normalized to the initial testbed with no changes, we see an increase in the power deposited in the treatment region as the dielectric property decreases. There is a decrease in the amount of power that reaches the target cell at large percentage media changes due to the dielectric mismatch at the border of the treatment region with the surrounding media. Despite that, the power in the full treatment region is not significantly different, indicating that in this case there may be no need to update the beamformer weights from the static weights determined prior to heating.  
	\begin{figure}[t!]
		\centering
		\includegraphics[width=1\linewidth]{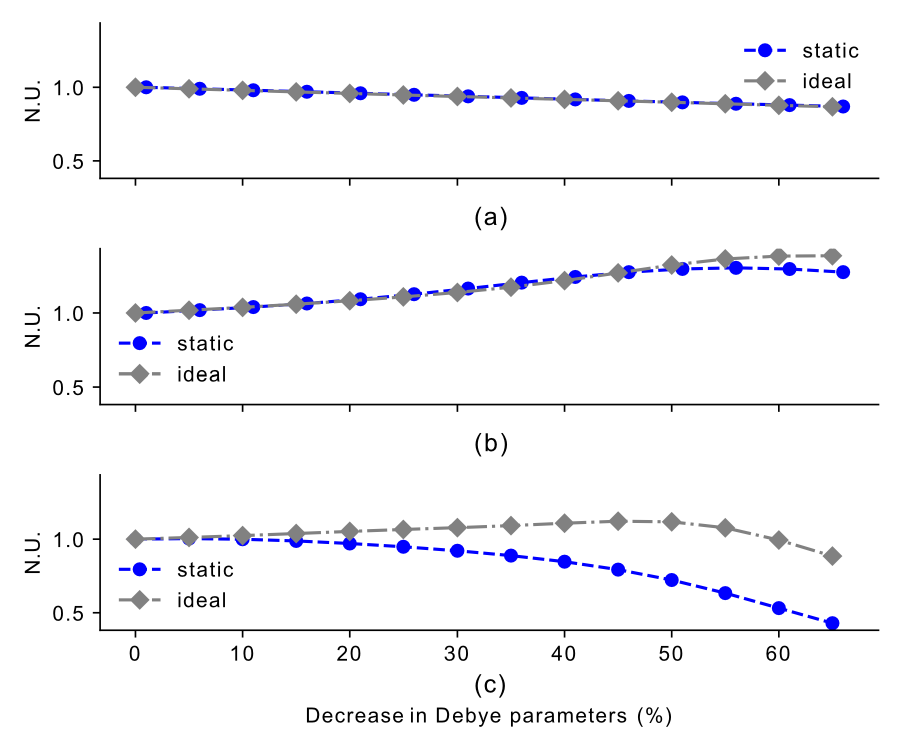}
		\caption{{Absorbed power by the (a) total media (b) treatment region and (c) target cell for the ideal and static beamformer when the Debye parameters of the treatment region is reduced \SI{5}{\percent} and the surrounding region is reduced by a factor of \SI{2}{\percent}.}}
		\label{fig:pwr_analysis_ablation_surround} 
	\end{figure}
	
	When electromagnetic waves travel through a highly dispersive media like fibroglandular tissue there will be power absorbed along the path to the target leading to material property changes in not just the treatment zone but in the surrounding region as well. To imitate a scenario where the propagating waves induce such changes in the surrounding region as well as the treatment zone we developed a model where Debye parameters of the treatment region are again changed by \SI{5}{\percent} up to \SI{35}{\percent}, while that of the surrounding region is changed by \SI{2}{\percent} up to \SI{74}{\percent}. The power deposition data is shown in Fig.~\ref{fig:pwr_analysis_ablation_surround}. As the surrounding region changes along with the treatment region, there is a drop in the overall power deposition in media for both the ideal and the static beamformer.  Decreasing the properties of the surrounding regions at different rates than the treatment region, there is a slight increase in the amount of power going into the treatment regions, even for the static beamformer. This is due to the fact that more power can now travel through the channel from the antennas furthest from the treatment region to the target. But, from \ref{fig:pwr_analysis_ablation_surround}(c), it is clear that the ideal beamformer is able to maintain focus at the target cell while for the static beamformer the deposited power reduces to more than half of the baseline.
	\begin{figure}[t!]
		\centering
		\includegraphics[width=1\linewidth]{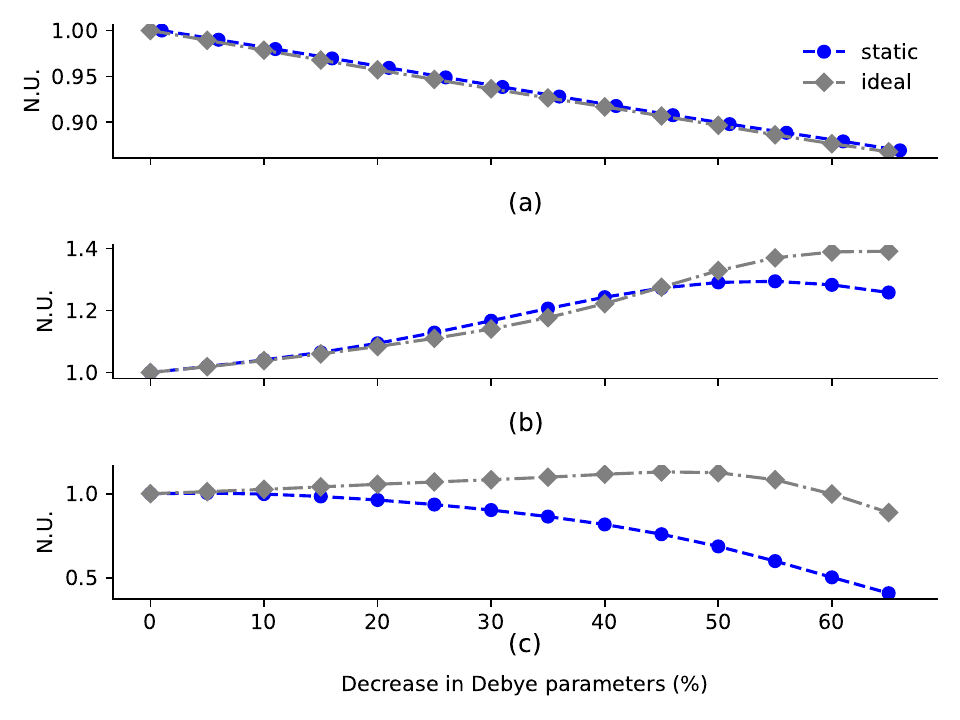}
		\caption{{Absorbed power by the (a) total media (b) treatment region and (c) target cell for the ideal and static beamformer when the treatment region is again reduced by  \SI{5}{\percent} and the surrounding region is reduced by a factor of \SI{2}{\percent} along with randomly placed hotspots which are reduced by \SI{1}{\percent}, \SI{3}{\percent} and \SI{4}{\percent} respectively.}}
		\label{fig:pwr_analysis_ablation_surround_hz_no_partial} 
	\end{figure}
	Fig. \ref{fig:pwr_analysis_ablation_surround_hz_no_partial} shows the same trends for the beamformers as Fig. \ref{fig:pwr_analysis_ablation_surround} even in the presence of three randomly located hotspots, showing similar results. 
	
	\subsection{Partial Knowledge Modeling}
	\label{sub_sec:obj_model}
	In this section we consider an adaptive beamformer that has access to only partial knowledge of the dielectric changes in the media, specifically an estimate of the bulk (average) dielectric changes. This models a real-time beamforming system that can more feasibly acquire average dielectric data than full dielectric maps, which would be too time consuming. In a practical setting the average dielectric changes could be estimated using various sensing approaches, such as those explored in~\cite{haldes2024estimating, haldes2025thesis}. In our simulation, we modeled this by scaling the Debye parameters by the previously defined percentages, averaged them and then calculated the average complex permittivity of the tissue using \eqref{eq:debye}. We then used that averaged dielectric data to update the beamformer weights by running the FDTD simulation. This process was repeated for each discrete iteration of the change in the dielectric property, following which the expected power deposition was calculated in the simulation.
	Note that since the temperature rise is not an instantaneous process and also depends on the amount of power being transmitted by the beamformer, the aforementioned parameters can be easily optimized to be compatible with real-time application. Here, we evaluate the performance of the beamformer relying on partial knowledge relative to the static and ideal beamformer.
	
\begin{figure}[t!]
		\centering
		\includegraphics[width=1\linewidth]{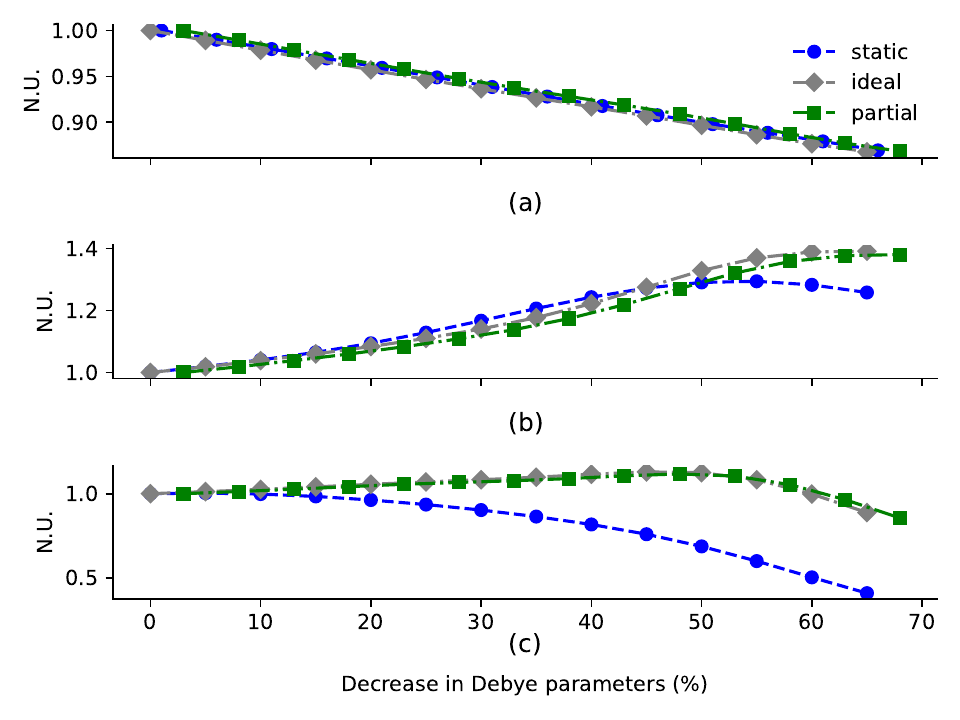}
		\caption{{Absorbed power by the (a) total media, (b) treatment region and (c) target cell for the ideal, static and the partial knowledge beamformer for the heterogeneous model. The treatment region is again reduced by  \SI{5}{\percent} and the surrounding region is reduced by a factor of \SI{2}{\percent} along with randomly placed hotspots which are reduced by \SI{1}{\percent}, \SI{3}{\percent} and \SI{4}{\percent} respectively.}}
		\label{fig:pwr_analysis_ablation_with_hotzones} 
	\end{figure}

\begin{figure}[t!]
		\centering
		\includegraphics[width=1\linewidth]{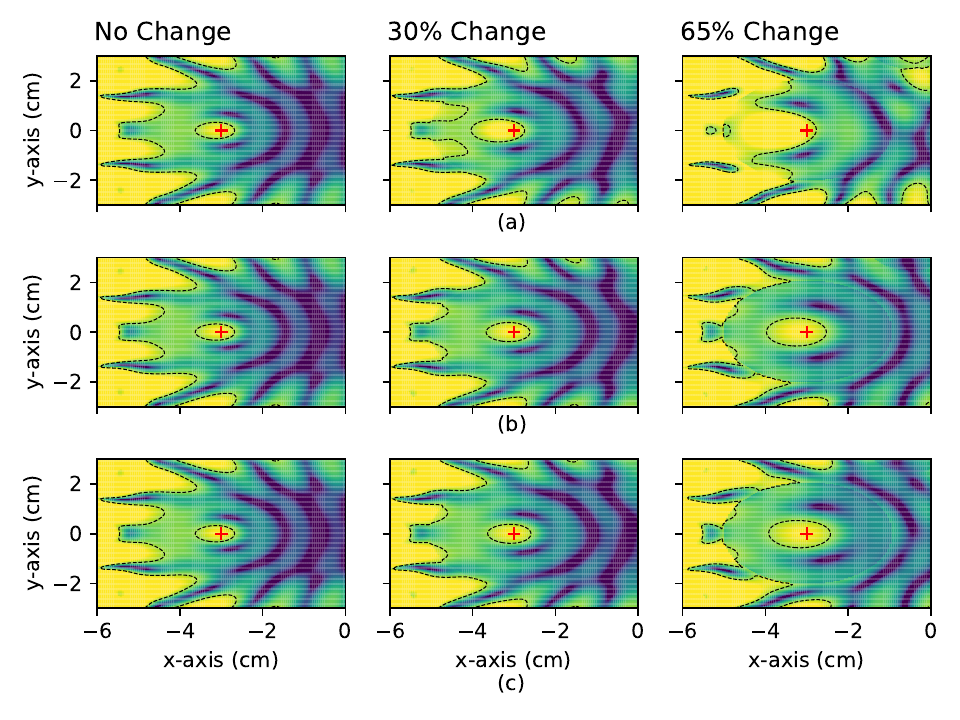}
		\caption{{Power deposition with \SI{-3}{\decibel} contouring for (a) static, (b) ideal and (c) partial knowledge beamformer. The baseline for all three cases were the homogeneous state with no changes in any regions. The middle column for all three cases indicate a \SI{30}{\percent} reduction in the treatment region Debye parameters. The surrounding region, hot zone 1, hot zone 2 and hot zone 3, experienced reductions of \SI{12}{\percent}, \SI{7}{\percent}, \SI{18}{\percent} and \SI{24}{\percent}, from the homogeneous state respectively. The third column shows the focusing accuracy of the beamformer when the treatment region Debye parameters are reduced to \SI{65}{\percent} of the homogeneous state. The surrounding region and the hot zones are similarly reduced by \SI{26}{\percent}, \SI{13}{\percent}, \SI{39}{\percent} and \SI{52}{\percent} respectively.}}
		\label{fig:2d_pwr_analysis_partial_knowledge} 
		\vspace{-\baselineskip}
	\end{figure}

	The model is given by Fig.~\ref{fig:pwr_analysis_ablation_with_hotzones}. 
	The trends shown by the results in Fig.~\ref{fig:pwr_analysis_ablation_with_hotzones} show that the power deposited in the total tissue region is almost the  same for all three cases. The power deposited in the treatment region and at the target cell, however, show appreciable differences between the static beamformer and the other two cases. Notably, the partial-knowledge beamformer and the ideal beamformer perform almost identically, showing minimal differences. This indicates that with only knowledge of the bulk changes in the material parameters, minimal loss of beamforming efficacy relative to an ideal beamformer is possible. Power deposition maps of the three cases are shown around $\pm$\SI{3}{\centi\meter} of the target in Fig.~\ref{fig:2d_pwr_analysis_partial_knowledge}. The power deposition has been contoured at \SI{-3}{\decibel} and as before the power deposition is normalized by the initial homogeneous baseline instance. In the figure the "No Change" case refers to the baseline for all three beamformer types.    
	{Two more cases from Fig.~\ref{fig:pwr_analysis_ablation_with_hotzones} were selected to highlight the gradual contrast of the performances of the three beamformers. The middle column in Fig.~\ref{fig:2d_pwr_analysis_partial_knowledge} shows the case when the Debye parameters of the treatment region are reduced by \SI{30}{\percent} while the surrounding areas and hot zones see reductions of \SI{12}{\percent}, \SI{7}{\percent}, \SI{18}{\percent} and \SI{24}{\percent}. The third column of the same figure represents a more extreme scenario where the treatment region parameters have been reduced by \SI{65}{\percent} of the homogeneous state with corresponding reductions in surrounding areas and hot zones of \SI{26}{\percent}, \SI{13}{\percent}, \SI{39}{\percent} and \SI{52}{\percent}.} 
	It can be seen that the focusing of the static beamformer gradually degrades completely, whereas both the partial knowledge and the ideal beamformer are able to maintain a focus. 
	
	In each case, the Debye parameters were reduced by the stated varying percentages. In the case of the initial 'no change' case, the calculated relative permittivity was \SI{47.43}{}. In the last case with \SI{65}{\percent} change, the calculated permittivity was \SI{33.20}{} at the ablation region. This accounts for about a \SI{30}{\percent} reduction in the actual relative permittivity of the region. When the spatial average of the of the tissue region is taken the relative permittivity was calculated as \SI{21.31}{}, which is about a \SI{55}{\percent} reduction from the ideal case. Despite these changes in the parameters, the partial knowledge beamforming approach maintains focus within the treatment region.
	\section{Multi-Objective Beamforming}
	\label{sec:mo_bf}
		In the above section we showed that for homogeneous media or a temperature independent heterogeneous test media the performance of the static beamformer is similar to that of an ideal beamformer. But, tissues where tumors are usually found, like in fibroglandular tissue or brain matter surrounded by cerebrospinal fluid, are highly dispersive tissue and their dielectric properties change with temperature due to absorption of microwave energy. Thus, we analyzed the effectiveness of the beamformer designs for a heterogeneous media and found that the static beamformer loses focus and causes unwanted power deposition in areas outside the target region. Furthermore, for both the ideal and the partial knowledge beamformer cases, while focusing is maintained, there is still unwanted power deposition in regions outside the treatment zone. This necessitates the need for a multi-objective beamformer design that can simultaneously heat cancerous cells while minimizing hotspots in healthy tissue regions. 
		
		
	\subsection{Simple Model}
	\label{sub_sec:mo_simple_test}
\begin{figure}
		\centering
		\includegraphics[width=1\linewidth]{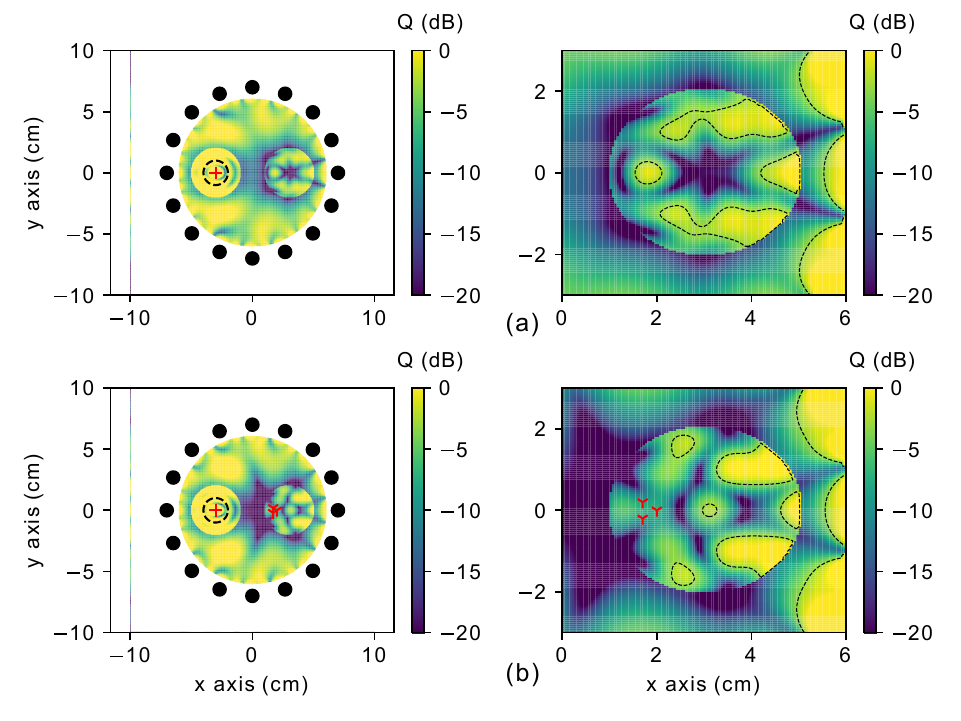}
		\caption{Comparison between (a) single objective and (b) multi-objective adaptive beamformer power deposition at the primary and secondary fibroglandular tissue inclusions surrounded by fatty tissue. The images on the right are contoured at -\SI{3}{\decibel}.}
		\label{fig:so_mo_sar_both}
		\vspace{-\baselineskip}
	\end{figure}

	We start with the simple model that has two regions of fibroglandular tissue inclusions of radius \SI{2}{\centi\meter} within fatty tissue. The spatial averaging adaptive approach is applied to the test media and the FDTD simulation is run to acquire the beamformer weights that perform single objective beamforming. In particular, we aim to steer a focus in the target region (the left region) and steer nulls at a hotspot in the other region (on the right). Fig.~\ref{fig:so_mo_sar_both} shows the beamformer performance. The power deposition data is normalized by the power deposited at the target which is used as the baseline case. The image on the right presents a magnified view of the secondary inclusion region showing absorbed power and is contoured at -\SI{3}{\decibel}. 
	Two cases are shown: a single-objective beamformer where only a focus is steered to the target region, and a multi-objective case where nulls are placed on a hotspot within the secondary region. It can be seen that placing three nulls in the multi-objective design mitigates one of the hotspots in the secondary region. Multiple nulls were used because null widths tend to be narrower than that of focii.
	The effectiveness of the nulls is evident, but there is still some power that is being absorbed by the tissue. To completely dissipate the power more nulls would be needed. In general, an $N$-element array can steer $N-1$ objectives, thus there is considerable leeway to mitigate hotspots further. In this work we aim to demonstrate the basic capability; a more detailed optimization study could be implemented to more thoroughly assess the mitigation performance. We also observe reduced power deposition in the surrounding fatty tissue region for the multi-objective beamformer. The higher levels of power absorption in the fatty tissue region around the fibroglandular tissue inclusions is mostly due to the abrupt change of dielectric properties. In a real patient the changes between the tissue materials would likely be more gradual.    
	\begin{figure}
		\centering
		\includegraphics[width=1\linewidth]{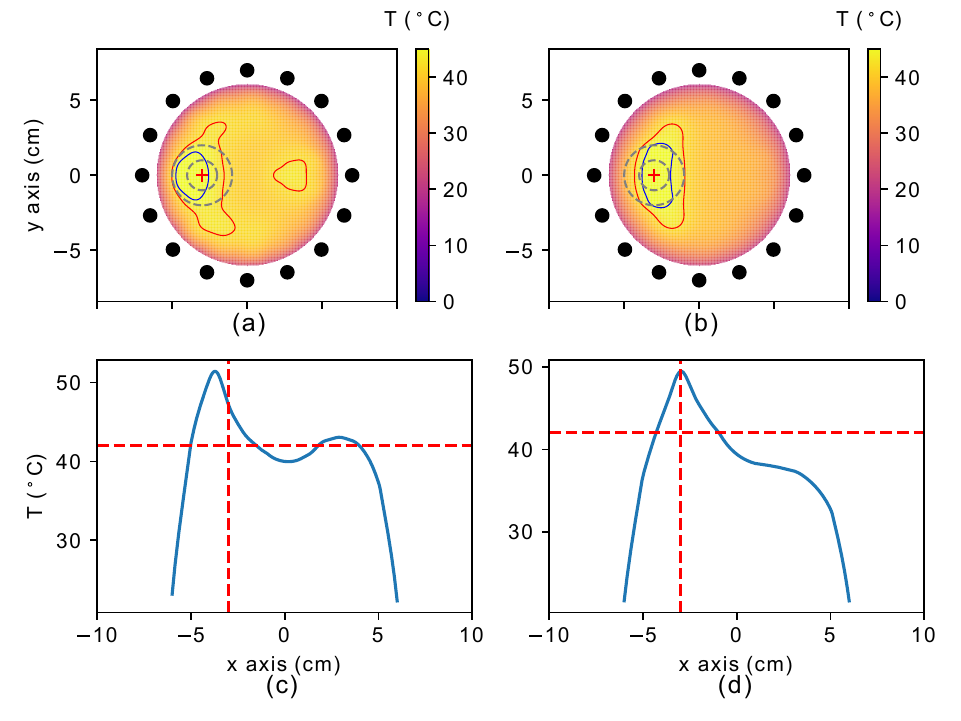}
		\caption{Thermal map of test environment for (a) single objective and (b) multi-objective adaptive beamformer. The thermal maps are contoured at threshold temperature \SI{42}{\degreeCelsius} and ablation temperature \SI{45}{\degreeCelsius}. The target region of \SI{1}{\centi \meter} and the treatment region of \SI{2}{\centi \meter} radius around the focus is denoted by the dashed lines on the 2D thermal maps. The temperature distribution is graphically represented by taking a 1D slice of the thermal map along the x-axis at $y = 0$ for (c) single objective and (b) multi-objective beamformer. The vertical red line depicts the location of the target cell and the horizontal red line is for threshold temperature the \SI{42}{\degreeCelsius}.}
		\label{fig:so_mo_thermal_both}
		\vspace{-\baselineskip}
	\end{figure}
	
	Importantly, higher power deposition levels do not necessarily translate to higher temperature regions. To fully validate the performance of the multi-objective beamformer it is necessary to analyze the temperature distribution of the test media for the two types of the beamformer. The power deposition data for both the beamformers were scaled by the same quantity to calculate the heating potential to generate the thermal maps as depicted in Fig.~\ref{fig:so_mo_thermal_both}. Fig.~\ref{fig:so_mo_thermal_both}(a) shows that for the single objective beamformer there is unwanted heating not only at the secondary inclusion region but also at the fatty tissue around the primary focus, which heats up to a temperature $\sim$ \SI{42}{\degreeCelsius}, depicted by the red contour line. The blue line in (a) shows the area heated to cell apoptosis temperatures of $\sim$ \SI{45}{\degreeCelsius}. The multi-objective beamformer significantly reduces auxiliary heating and actively concentrates the primary focus within the treatment region. A 1D cut of the temperature profile is taken along the x-axis at $y = 0$ and shown in Figs~\ref{fig:so_mo_thermal_both}(c) and (d).
	Fig. \ref{fig:so_mo_thermal_both}(c) shows that for the adaptive single objective beamformer the peak temperature is slightly shifted off the target location. The multi-objective beamformer corrects for the shift while lowering the temperature at the unwanted hotspot. 
	
	The observed high temperatures in the fatty tissue surrounding primary fibroglandular inclusion can be largely attributed to the sharp contrasts in the thermal and dielectric properties. Steering the nulls helps reduce the temperatures at the secondary inclusion but not completely in the primary inclusion region. Additional nulls could be placed in the region directly adjacent to the boundary between the two tissues to reduce the temperatures however that may not be necessary in a more realistic scenario.
	Note that with a detailed initial dielectric map, areas where hotspots manifest can be predicted, and thus nulls can be designed into the initial multi-objective beamformer. 
	This presents an added advantage for the non-invasive system as the algorithm is mostly independent of feedback from the heterogeneous test media, and can increase the time interval between updating the beamformer weights to maintain focusing. 
	The multi-objective beamformer could also be used to steer multiple focii to multiple treatment regions, steering a total of $M = N-1$ focii or nulls.
	
	\subsection{Realistic Phantom}
	\label{sub_sec:mri_model}
	
\begin{figure}
		\centering
		\includegraphics[width=1\linewidth]{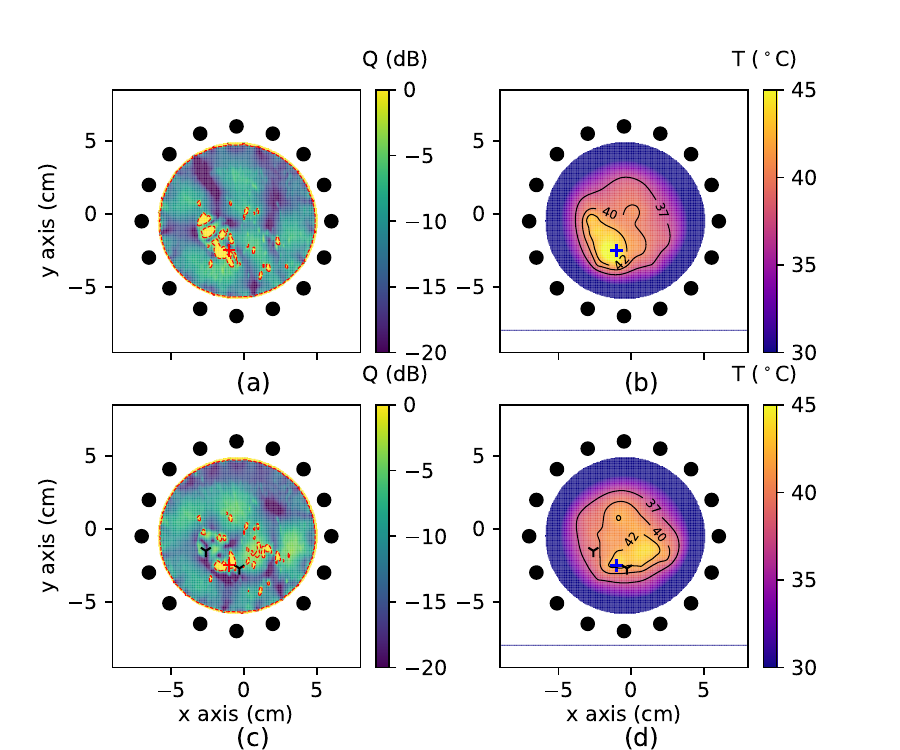}
		\caption{Virtual patient model analysis: (a) single objective power deposition, (b) thermal image of single objective beamforming with conturing at \SI{37}{}, \SI{40}{} and \SI{42}{\degreeCelsius}. (c) Power deposition image for multi-objective beamforming with one focus and two nulls placed in the regions of unwanted heating. (d) Thermal distribution map with the same conturing as (b).}
		\label{fig:em_thermal_vp_so_mo}
	\end{figure}

	In this section we evaluate the partially adaptive non-invasive multi-objective beamformer for a realistic virtual patient. We chose the class II MRI-derived breast phantom with scattered fibroglandular tissue immersed in water as described in Section~\ref{sub_sec:bf_setup} with the dielectric map given in Fig.~\ref{fig:virtual_patient}. The power deposition of the single objective beamformer is shown in Fig.~\ref{fig:em_thermal_vp_so_mo}(a) depicting unwanted power deposition with the scattered fibroglandular tissue at locations other than the target. However, the corresponding thermal map in Fig.~\ref{fig:em_thermal_vp_so_mo}(b) shows that not all locations with high power deposition are appreciably heated. Furthermore, there is no edge heating due to the better impedance matching between skin and the water, and the active cooling effect of the surrounding water, helping to maintain the tissue temperature at $\sim$ \SI{37}{\degreeCelsius}. The area immediately around the target region reaches the temperature $\sim$ \SI{45}{\degreeCelsius} as shown by the contouring. This unwanted heating can be prevented by utilizing the multi-objective beamforming LCMP algorithm.
	Both the power deposition and the thermal map in Fig.~\ref{fig:em_thermal_vp_so_mo}(c) and (d) show that by placing the two nulls in the regions most affected, the unwanted heating can be mitigated. It is important to note here that the nulls are very narrow and thus some of the power can migrate to a different region causing a new hot spot. An effective strategy to mitigate unwanted heating for models that have more percentage of fibroglandular tissue would be to have more than one null be scanned over probable hot zones as a function of time to attempt to maintain thermal equilibrium over other regions. 
Alternatively, if the model is mostly fat or media with similar dielectric properties, nulls may not be required at all. 

\section{Result Analysis}
To quantify the precision and the selectivity of our beamformer design we use the following two metrics, which are based on those used in \cite{zastrow20103d}:

\begin{itemize}
	\item[1.] Accuracy: the radial distance from the peak temperature location to the intended target location inside the tissue model under observation.
	
	\item[2.] Thermal Selectivity ($\mathbf{A_{45}}$): the area of the tissue around the target region that reaches temperatures $\geq$ \SI{45}{\degreeCelsius}
\end{itemize}
For the evaluation we used the homogeneous fibroglandular model, the simple heterogenous testbed from section \ref{sub_sec:mo_simple_test} and the realistic breast phantom from section \ref{sub_sec:mri_model}. We compared the performance of both the single objective beamformer and the multi-objective beamformer for the heterogeneous model and the virtual patient model.  
\begin{table}[tb]
	\ifcsname hlon\endcsname%
	\color{blue}%
	\fi%
	\caption{{Accuracy and Resolution}}
	\label{tab:result_analysis}
	\begin{center}
		\begin{tabularx}{\columnwidth}{p{0.26\linewidth}p{0.15\linewidth}YYY}
			\toprule[1pt]
			\midrule
			Tissue Type & Temperature at Target [\SI{}{\degreeCelsius}] & Maximum Temperature [\SI{}{\degreeCelsius}]  & Accuracy [\SI{}{\milli\meter}] & $\mathbf{A_{45}}$ [\SI{}{\centi\meter}$^2$] \\
			\midrule
			Homogeneous Fibroglandular & \SI{41.9}{} & \SI{42.04}{} & \SI{18.00}{} & 1.79\\[0.5ex]
			\midrule
			Simple Heterogeneous (Single Objective) [\ref{sub_sec:mo_simple_test}]  & \SI{47.12}{}  & \SI{51.42}{} & \SI{6.9}{} & \SI{5.29}{} \\ [0.5ex]
			\midrule
			Simple Heterogeneous (Multi-Objective) [\ref{sub_sec:mo_simple_test}] & \SI{49.47}{} & \SI{49.46}{} & \SI{0.49}{} & \SI{2.45}{} \\ [0.5ex]
			\midrule
			Realistic Phantom (Single Objective) [\ref{sub_sec:mri_model}] & \SI{46.46}{} & \SI{47.76}{} & \SI{3.0}{} & \SI{1.05}{} \\[0.5ex]
			\midrule
			Realistic Phantom (Multi-Objective) [\ref{sub_sec:mri_model}] & \SI{40.46}{} & \SI{43.89}{} & \SI{4.1}{} & \SI{0.87}{} \\[0.5ex]
			\midrule
			\bottomrule[1pt]	
		\end{tabularx}
	\end{center}
\end{table}
Table \ref{tab:result_analysis} lists the temperature at the target, the peak temperature within the tissue, the accuracy, and thermal selectivity for all the test models. The peak temperature for all the models is within the target region of \SI{1}{\centi\meter} except for the simple homogeneous testbed. 
The heterogeneous models are phantoms containing fibroglandular and fatty tissue. Fatty tissue absorbs less energy, resulting in greater power deposition at the target for testbeds that have fatty tissue, and a higher accuracy. In contrast, the homogeneous testbed is entirely composed of fibroglandular tissue, which absorbs more energy. This results in lower accuracy, a larger area around the target gets heated to ablation temperatures. As intended, the multi-objective beamformer helps increase selectivity when compared to the single objective for both heterogeneous testbeds under consideration.{The complex environment of the realistic breast phantom induces more spatial changes in the power distribution in the multi-objective case compared to the single objective case, causing slightly increased heating in some areas and slightly decreased heating at the target region. As a result, the target reaches a temperature of \SI{40.46}{\degreeCelsius} for the multi-objective beamforming. This may be mitigated by scanning multiple nulls over time throughout the tissue region. 
	
	\section{Conclusion}
	
	We evaluated the feasibility of a multi-objective microwave beamformer for heating tissues in thermally dynamic environments for non-invasive microwave hyperthermia.
	In particular, this work aimed to develop a generalized beamformer capable of adapting to real-time changes in the propagation channel using only estimates of the average change in the dielectric properties of the media. The goal was to achieve cell apoptosis temperatures ($\sim \SI{45}{\degreeCelsius}$) through selective electromagnetic wave absorption and focusing, with minimal reliance on instantaneous knowledge of the dielectric media variations, while simultaneously creating nulls to prevent off-target heating.
	We demonstrated the efficacy of a multi-objective adaptive beamformer, starting with simplified homogeneous media in air and water to establish crucial physical system design parameters. Our analysis indicated that for a \SI{12}{\centi \meter} diameter observation area, a 16-element array offered an optimal balance between focus precision and practical implementation.
	A key advantage of this approach lies in its inherent flexibility in placing nulls. Unlike many existing methods available in literature that depend on feedback from the treatment media in the form of SAR or thermal maps – a major obstacle for non-invasive systems – our design leverages an initial MRI-derived dielectric map that can be used for hotspot prediction. By strategically positioning nulls towards potential areas of concern, such as secondary fibroglandular inclusions or interfaces between different tissue types, unwanted heating can be proactively mitigated. This capability offers a substantial benefit for non-invasive applications by reducing the need for frequent channel weight acquisition while still maintaining effective focusing. This not only simplifies system implementation and potentially lowers computational demands but also enhances the system's resilience to variations in tissue properties. The study presented here is based on 2D modeling and does not account for power deposition and heating in the longitudinal axis. Future work will evaluate the full effect of the beamformer design though a comprehensive study on the computational requirements, impacts of errors such as dielectric averaging estimates, and a full 3D model analysis.
	
	\bibliography{biblatex_all.bib}
	\bibliographystyle{IEEEtran}
	
	\begin{IEEEbiography}[{\includegraphics[width=1in,height=1.25in,clip,keepaspectratio]{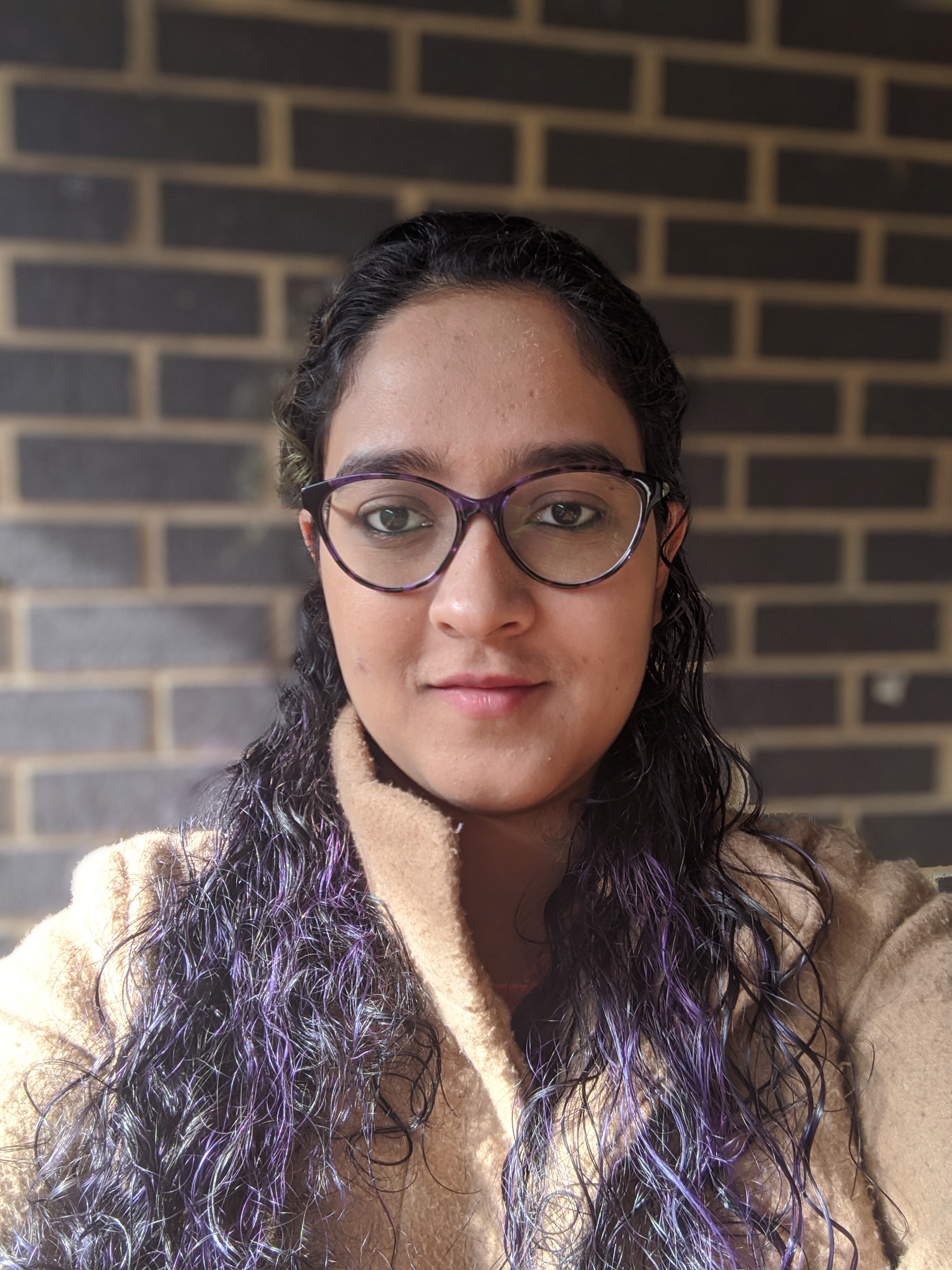}}]{Ahona Bhattacharyya} (Graduate Student Member, IEEE) received the B.Tech. degree in electronics and communication engineering from, Maulana Abul Kalam Azad University of Technology, Kolkata, India and M.S. degree from University at Buffalo, State University of New York, in 2018 and 2020, respectively.

In 2020, she joined the Department of Electrical and Computer Engineering, Michigan State University, East Lansing, MI, USA, where she is presently a graduate research assistant. Her current research interests include distributed arrays, near-field adaptive beamforming, radar and signal processing, microwave wireless systems, and electromagnetic material characterization.
\end{IEEEbiography}
\vskip -2\baselineskip plus -1fil
\begin{IEEEbiography}[{\includegraphics[width=1in,height=1.25in,clip,keepaspectratio]{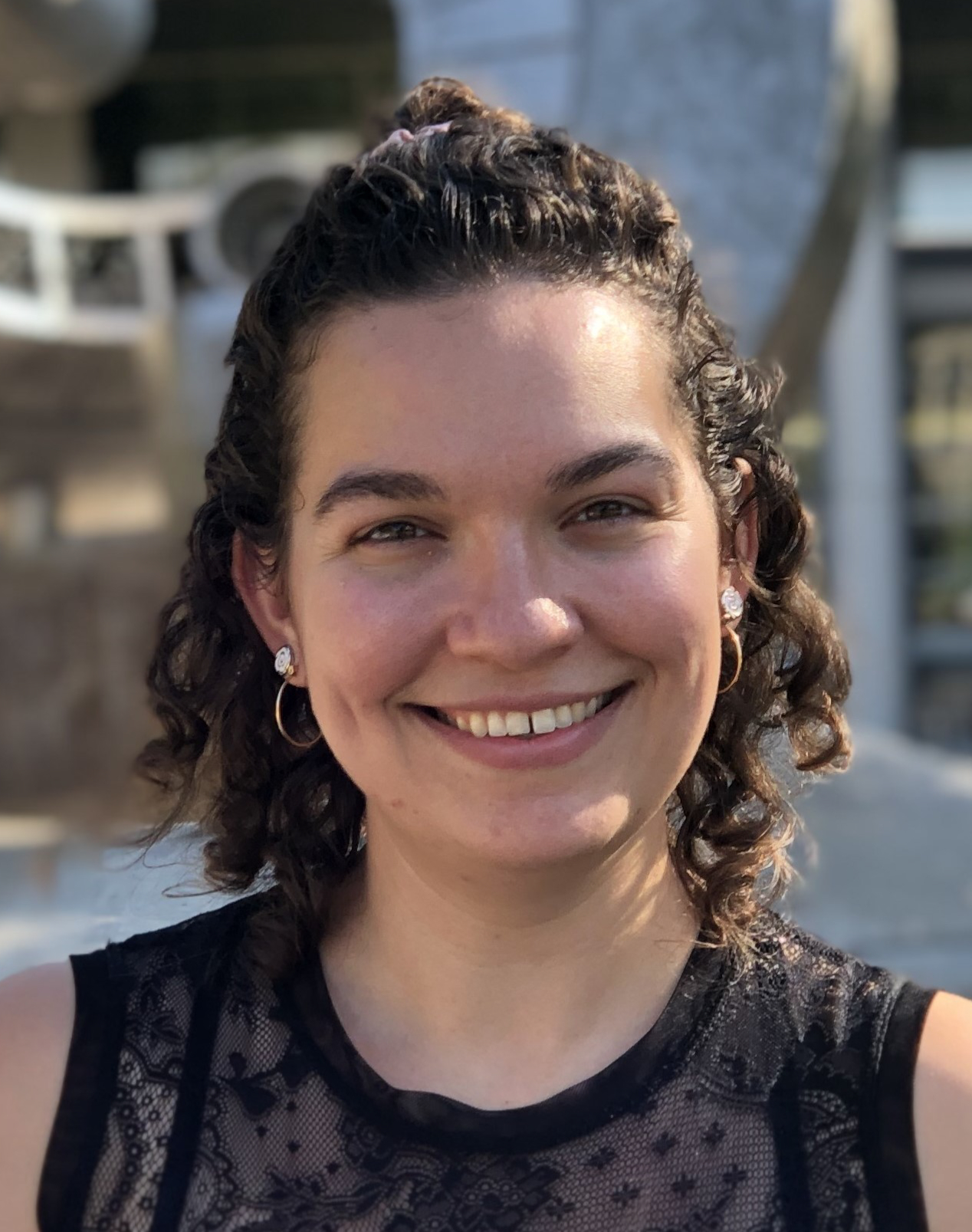}}]{Tessa A. Haldes} (Member, IEEE) received the B.S. degree in electrical engineering from Northwestern University, Evanston, IL, USA, in 2016 and the M.S. and Ph.D. degrees in electrical engineering from the University of Wisconsin-Madison, Madison, WI, USA, in 2021 and 2025, respectively.
	
She is currently a Senior Antenna Design Engineer at NuCurrent, Chicago, IL, USA, where she develops electrically small antennas for wireless power transfer applications, including consumer electronics and wearable and implantable medical devices. Prior to pursuing graduate studies in 2019, she worked as an Electrical Engineering Contractor with a primary focus on medical device development.
	
Her research interests include applications of electromagnetics in medicine, particularly the interactions between electromagnetic waves and biological tissue for imaging, diagnostics, and treatment, as well as the development of novel wireless power transfer solutions. Dr. Haldes was the recipient of the 2022 IEEE Antennas and Propagation Society (AP-S) Doctoral Research Grant.
\end{IEEEbiography}
\vskip -3.2\baselineskip plus -1fil
\begin{IEEEbiography}[{\includegraphics[width=1in,height=1.25in,clip,keepaspectratio]{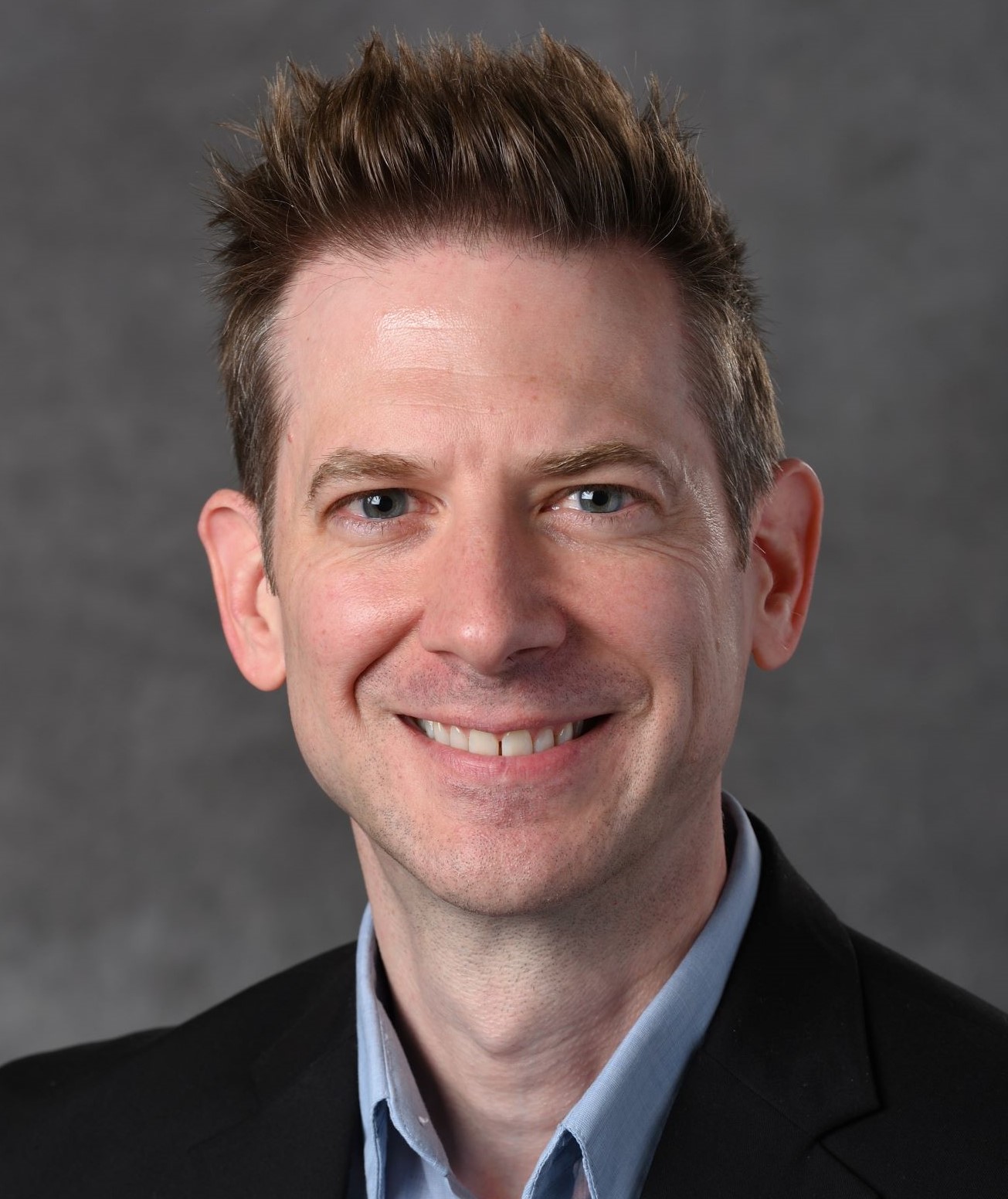}}]{Jeffrey A. Nanzer} (Senior Member, IEEE) received the B.S. degrees in electrical engineering and in computer engineering from Michigan State University, East Lansing, MI, USA, in 2003, and the M.S. and Ph.D. degrees in electrical engineering from The University of Texas at Austin, Austin, TX, USA, in 2005 and 2008, respectively.
	
From 2008 to 2009 he was with the University of Texas Applied Research Laboratories in Austin, Texas as a Post-Doctoral Fellow designing electrically small HF antennas and communications systems. From 2009 to 2016 he was with the Johns Hopkins University Applied Physics Laboratory where he created and led the Advanced Microwave and Millimeter-Wave Technology Section. In 2016 he joined the Department of Electrical and Computer Engineering at Michigan State University where he held the Dennis P. Nyquist Assistant Professorship from 2016 through 2021. He is currently a Professor. He directs the Electromagnetics Laboratory, which consists of the Antenna Laboratory, the Radar Laboratory, and the Wireless Laboratory. He has published more than 250 refereed journal and conference papers, two book chapters, and the book Microwave and Millimeter-Wave Remote Sensing for Security Applications (Artech House, 2012). His research interests are in the areas of distributed phased arrays, dynamic antenna arrays, millimeter-wave imaging, remote sensing, millimeter-wave photonics, and electromagnetics.
	
Dr. Nanzer is a member of the IEEE Antennas and Propagation Society Education Committee and the USNC/URSI Commission B. He was a founding member and the First Treasurer of the IEEE APS/MTT-S Central Texas Chapter, served as the Vice Chair for the IEEE Antenna Standards Committee from 2013 to 2015, and served as the Chair of the Microwave Systems Technical Committee (MTT-16), IEEE Microwave Theory and Techniques Society from 2016 to 2018. He was a Distinguished Microwave Lecturer for the IEEE Microwave Theory and Techniques Society (Tatsuo Itoh Class of 2022-2024) and a Guest Editor of the Special Issue on Special Issue on Radar and Microwave Sensor Systems in the IEEE Microwave and Wireless Components Letters in 2022. He was a recipient of the Withrow Junior Distinguished Scholar Award in 2024, the Google Research Scholar Award in 2022 and 2023, the IEEE MTT-S Outstanding Young Engineer Award in 2019, the DARPA Director’s Fellowship in 2019, the National Science Foundation (NSF) CAREER Award in 2018, the DARPA Young Faculty Award in 2017, and the JHU/APL Outstanding Professional Book Award in 2012. 
\end{IEEEbiography}
\vskip 0pt plus -1fil
\begin{IEEEbiography}[{\includegraphics[width=1in,height=1.25in,clip,keepaspectratio]{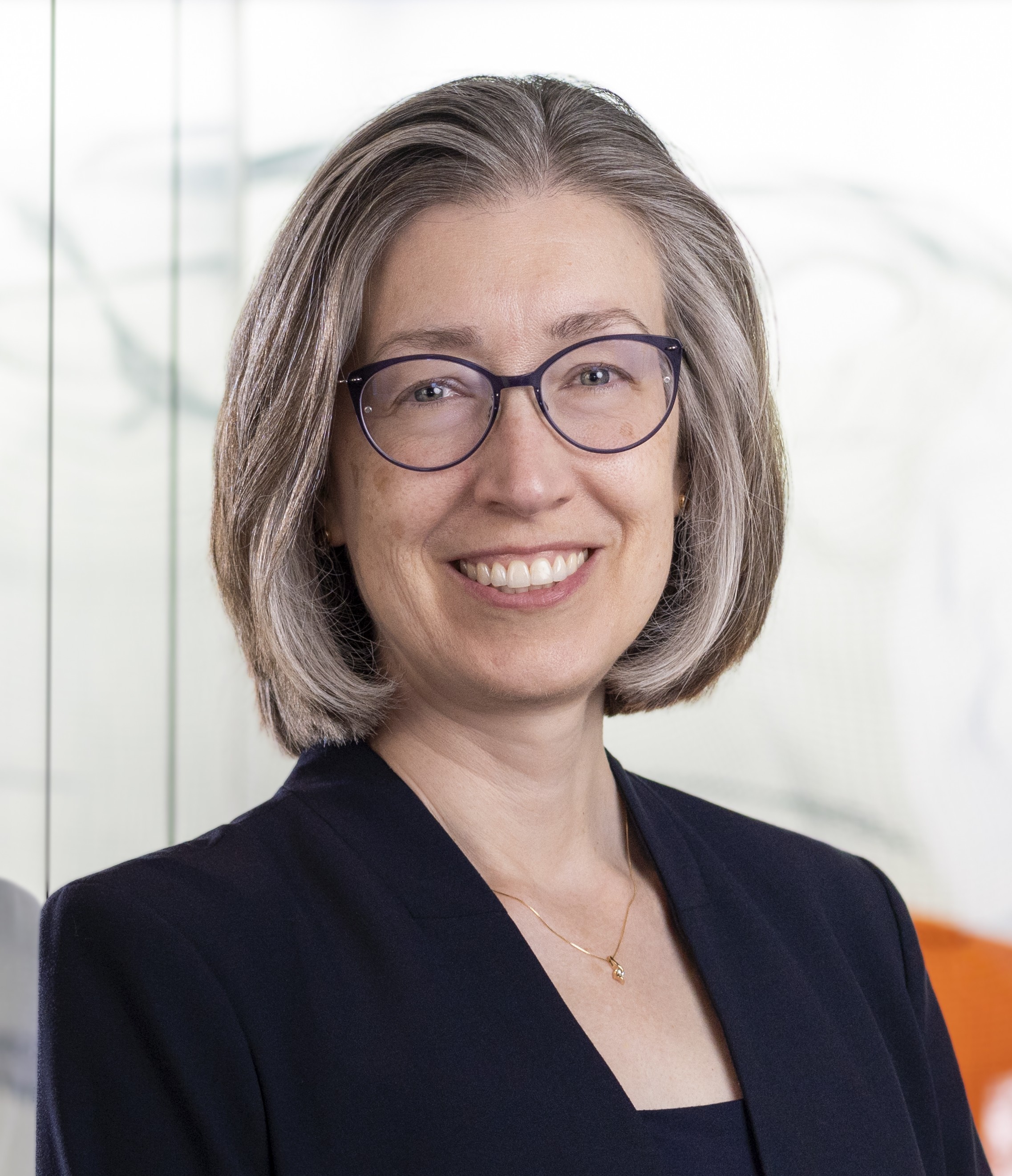}}]{Susan C. Hagness}(Fellow, IEEE) is the Charles H. Phipps Dean of the Case School of Engineering at Case Western Reserve University and a professor of the Department of Electrical, Computer, and Systems Engineering. She is also the Philip D. Reed Professor Emeritus and the Maria Stuchly Professor Emeritus at the University of Wisconsin-Madison where she served on the faculty for 28 years, including as the Associate Dean for Research and Graduate Affairs in the College of Engineering (2014-2017) and as the Chair of the Department of Electrical and Computer Engineering (2018-2026).

Dr. Hagness is a member of the National Academy of Engineering (NAE) and a Fellow of the IEEE, the American Association for the Advancement of Science (AAAS), the American Institute for Medical and Biological Engineering (AIMBE), and the National Academy of Inventors (NAI). She has held a variety of professional society appointments and elected leadership roles within the IEEE Antennas and Propagation Society, the IEEE Engineering in Medicine and Biology Society, and the U.S. National Committee (USNC) of the International Union of Radio Science (URSI).

She was the recipient of the Presidential Early Career Award for Scientists and Engineers (PECASE) presented by the White House in 2000. In 2002, she was named one of the 100 top young innovators in science and engineering in the world by the MIT Technology Review magazine. She is also the recipient of the UW-Madison Emil Steiger Distinguished Teaching Award (2003), the IEEE Engineering in Medicine and Biology Society Early Career Achievement Award (2004), the URSI Issac Koga Gold Medal (2005), the IEEE Transactions on Biomedical Engineering Outstanding Paper Award (2007), the IEEE Education Society Mac E. Van Valkenburg Early Career Teaching Award (2007), the UW System Alliant Energy Underkofler Excellence in Teaching Award (2009), the Physics in Medicine and Biology Citations Prize (2011), the Sven Berggren Prize from the Royal Physiographic Society of Lund, Sweden (2015), UW-Madison College of Engineering awards for teaching (2014), research (2018), and equity and diversity efforts (2021), the UW-Madison Women Faculty Mentoring Program Slesinger Award for Excellence in Mentoring (2017), UW-Madison Vice-Chancellor for Research honors including a Kellett Mid-Career award (2011) and a WARF Named Professorship (2024), the USNC-URSI Impact Award (2026), and the IEEE AP-S Distinguished Achievement Award (2026). She has graduated 30 Ph.D. students, published over 115 peer-reviewed journal papers, and is an inventor on 13 U.S. patents.
\end{IEEEbiography}

\end{document}